\documentclass{elsart}
\usepackage{ifpdf}
\usepackage{graphicx,amssymb,lineno}
\usepackage{multirow}
\ifpdf
\usepackage[%
  pdftitle={Design and performance of the muon monitor for the T2K neutrino oscillation experiment},%
  pdfauthor={Kodai Matsuoka},%
  pdfsubject={The preprint document class elsart},%
  pdfkeywords={instructions for use, elsart, document class},%
  pdfstartview=FitH,%
  bookmarks=true,%
  bookmarksopen=true,%
  breaklinks=true,%
  colorlinks=true,%
  linkcolor=blue,anchorcolor=blue,%
  citecolor=blue,filecolor=blue,%
  menucolor=blue,pagecolor=blue,%
  urlcolor=blue]{hyperref}
\else
\usepackage[%
  breaklinks=true,%
  colorlinks=true,%
  linkcolor=blue,anchorcolor=blue,%
  citecolor=blue,filecolor=blue,%
  menucolor=blue,pagecolor=blue,%
  urlcolor=blue]{hyperref}
\fi

\makeatletter
\def\elsartstyle{%
    \def\normalsize{\@setfontsize\normalsize\@xiipt{14.5}}
    \def\small{\@setfontsize\small\@xipt{13.6}}
    \let\footnotesize=\small
    \def\large{\@setfontsize\large\@xivpt{18}}
    \def\Large{\@setfontsize\Large\@xviipt{22}}
    \skip\@mpfootins = 18\p@ \@plus 2\p@
    \normalsize
}
\@ifundefined{square}{}{}
\makeatother

\begin{document}

\begin{frontmatter}
\title{Design and performance of the muon monitor for the T2K neutrino oscillation experiment}

\author[Kyoto]{K.~Matsuoka\corauthref{cor}},
\corauth[cor]{Corresponding author.}
\ead{matsuoka@scphys.kyoto-u.ac.jp}
\author[Kyoto]{A.K.~Ichikawa},
\author[Kyoto]{H.~Kubo},
\author[Osaka]{K.~Maeda},
\author[KEK]{T.~Maruyama},
\author[Osaka]{C.~Matsumura},
\author[Kyoto]{A.~Murakami},
\author[Kyoto]{T.~Nakaya},
\author[KEK]{K.~Nishikawa},
\author[Osaka]{T.~Ozaki},
\author[KEK]{K.~Sakashita},
\author[Kyoto]{K.~Suzuki},
\author[KEK]{S.Y.~Suzuki},
\author[Osaka]{K.~Tashiro},
\author[Osaka]{K.~Yamamoto},
\author[Tokyo]{M.~Yokoyama}
\address[Kyoto]{Department of Physics, Kyoto University, Kyoto 606-8502, Japan}
\address[Osaka]{Department of Physics, Osaka City University, Osaka 588-8585, Japan}
\address[KEK]{High Energy Accelerator Research Organization (KEK), Tsukuba 305-0801, Japan}
\address[Tokyo]{Department of Physics, University of Tokyo, Tokyo 113-0033, Japan}


\begin{abstract}
This article describes the design and performance of the muon monitor for the T2K (Tokai-to-Kamioka) long baseline neutrino oscillation experiment.
The muon monitor consists of two types of detector arrays: ionization chambers and silicon PIN photodiodes.
It measures the intensity and profile of muons produced, along with neutrinos, in the decay of pions.
The measurement is sensitive to the intensity and direction of the neutrino beam.
The linearity and stability of the detectors were measured in beam tests to be within 2.4\% and 1.5\%, respectively.
Based on the test results, the precision of the beam direction measured by the muon monitor is expected to be 0.25~mrad.
\end{abstract}

\end{frontmatter}

\section{Introduction}

The T2K (Tokai-to-Kamioka) experiment~\cite{T2K} is a long baseline neutrino oscillation experiment.
An intense muon neutrino beam is produced by using the \mbox{30-GeV} proton beam at \mbox{J-PARC} (Japan Proton Accelerator Research Complex), Tokai.
A train of eight~bunches of protons is extracted as a spill from the proton synchrotron main ring to the neutrino beamline.
The bunch width is 58~nsec, the bunch interval is 581~nsec, and the spill cycle is 3.5~seconds.
The proton beam is transported to and impinges on the graphite target to produce charged pions.
The pions are focused by three magnetic horns pulsed with 320~kA.
The pions enter the \mbox{96-m} decay volume and decay mainly into pairs of a muon and a muon neutrino.
The neutrinos before oscillation are measured by the near detector (ND280) at \mbox{J-PARC}.
After traveling 295~km, they are detected by the Super-Kamiokande detector~\cite{Super-K} (\mbox{Super-K}) in the Kamioka Observatory.

\begin{figure}[b]
  \begin{center}
    \includegraphics[keepaspectratio=true,width=140mm]{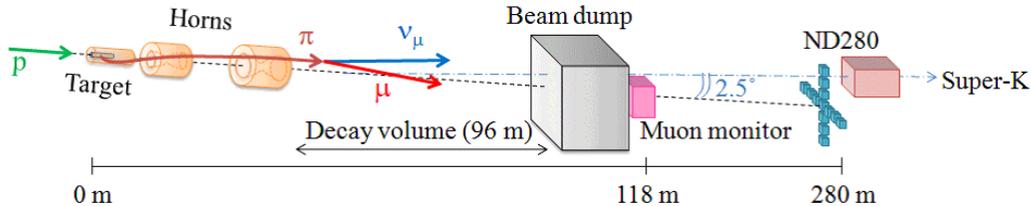}
  \end{center}
  \caption{Schematic view of the T2K neutrino beamline.
  The three horns and the muon monitor are aligned on the beam axis.
  An example trajectory of a pion decaying into a muon and a muon neutrino is also shown.}
  \label{fig:beamline}
\end{figure}

One of the novel aspects of the T2K experiment is that it has an off-axis beam configuration, that is, the neutrino beam is aimed at a direction away from \mbox{Super-K} by 2.5~degrees (Fig.~\ref{fig:beamline}).
This configuration makes the neutrino energy spectrum measured at \mbox{Super-K} narrower~\cite{off-axis} than one in a conventional on-axis beam configuration.
The narrow-band beam maximizes the sensitivity to the neutrino oscillation.
The neutrino beam energy, however, varies sensitively as a function of the off-axis angle.
Therefore, it is important to monitor and control the beam direction precisely; the beam direction is required to be aimed within 1~mrad.

To monitor the beam intensity and direction, the muon monitor is placed in the muon pit, which is just downstream of the beam dump and 18.5~m under the ground.
The distance from the proton target to the muon monitor is 118~m.
The muon monitor measures a position profile of muons produced along with neutrinos from the same parent pions.
The neutrino beam direction is measured as a direction from the proton target to the center of the muon profile.

A muon beam is also used to monitor the neutrino beam intensity and direction in other accelerator neutrino experiments~\cite{1967 MUMON, NOMAD MUMON, IHEP MUMON, AGS MUMON, K2K MUMON, CNGS MUMON}.
In the NuMI (Neutrinos at the Main Injector) neutrino beamline~\cite{NuMI} at Fermi National Accelerator Laboratory (FNAL), ionization chambers filled with pure He gas are used for muon monitors.
The NuMI ionization chamber~\cite{NuMI MUMON} was referred to on our ionization chamber design.

In Sec.~\ref{sec:requirement}, the requirements for the muon monitor are described with the properties of the T2K muon beam.
In Sec.~\ref{sec:instruments}, the design of the muon monitor detectors and instruments is described.
In Sec.~\ref{sec:beam test}, the performances of the detectors measured in beam tests are described.

\section{Requirements for the muon monitor}\label{sec:requirement}

\begin{figure}[pth]
    \begin{center}
      \includegraphics[keepaspectratio=true,width=80mm]{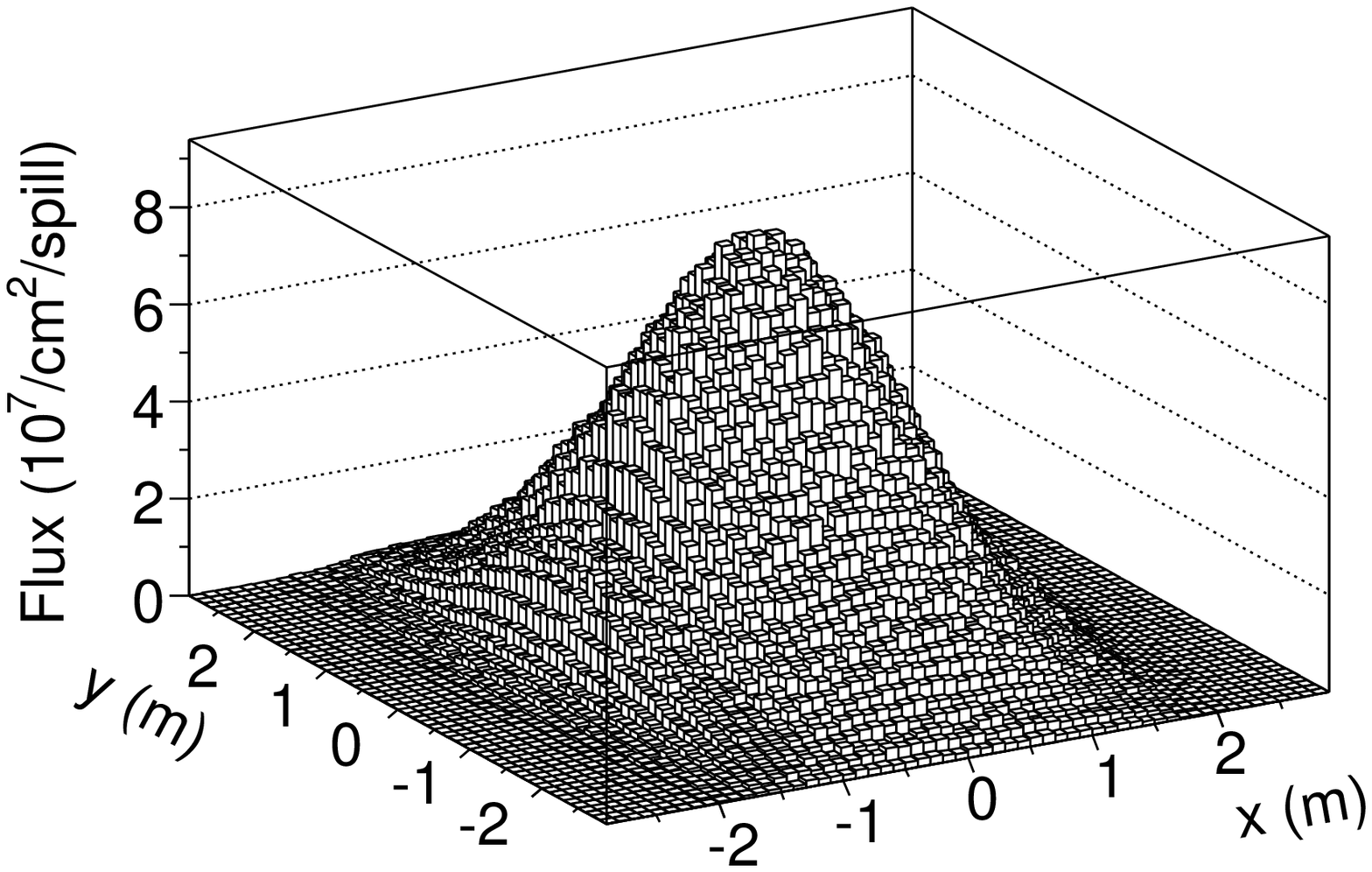}
    \end{center}
    \caption{Two-dimensional profile of charged particles at the muon monitor by the Monte Carlo simulation. The horizontal and vertical axes are denominated x and y, respectively.}
    \label{fig:MUMON profile}
  \vspace{20pt}

  \begin{center}
    \includegraphics[keepaspectratio=true,width=140mm]{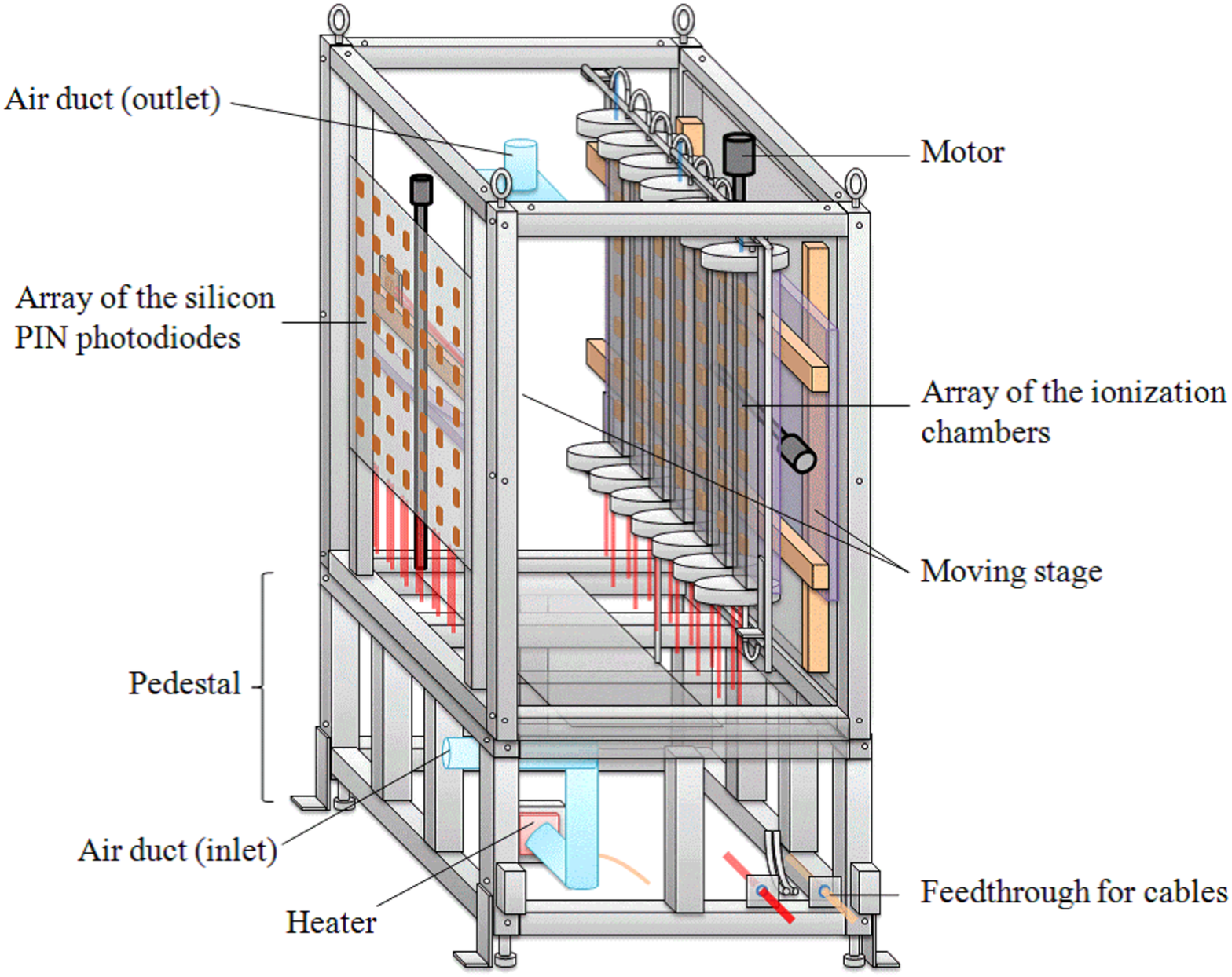}
  \end{center}
  \caption{Schematic view of the muon monitor.
  The beam enters from the left side.
  On the upstream side, 49~silicon PIN photodiodes are placed on the support enclosure.
  On the downstream side, the large moving stage holds seven ionization chambers, each of which contains seven~sensors.
  The whole structure is covered with aluminum insulation panels, which are not drawn in the figure.}
  \label{fig:MUMON}
\end{figure}

The requirements for the muon monitor are:
\begin{enumerate}
  \item to monitor the stability of the neutrino beam intensity with a precision better than 3\% by measuring the muon beam intensity;
  \item to measure the neutrino beam direction with a precision better than 0.25~mrad, which corresponds to a 3-cm precision of the muon profile center;
  \item to measure the beam intensity and direction on a bunch-by-bunch basis;
  \item to measure the intense muon beam of $10^5$-$10^7$~/cm$^2$/bunch;
  \item to be tolerant of radiation and resistant to activation;
  \item to run stably to minimize deadtime due to repair.
\end{enumerate}

A muon energy threshold for the beam dump is chosen to minimize the hadron flux at the muon monitor while keeping the sensitivity to the neutrino beam direction; only muons with energy above 5~GeV can go through the beam dump and reach the muon monitor.
An expected profile of charged particles in a Monte Carlo simulation is shown in Fig.~\ref{fig:MUMON profile}.
It is close to a Gaussian distribution with a sigma of approximately 1~m around the beam center.
To reconstruct the profile, the muon monitor is equipped with 49~sensors in a $150\times 150$~cm$^2$ plane.
The number of charged particles at the profile center is estimated at $8\times10^7$~/cm$^2$/spill for $3.3\times10^{14}$~protons/spill.
The expected ratio of the muons to all the charged particles is around 87\% and the remaining percentage is predominantly due to $\delta$-rays knocked out by the muons.
An absorbed dose at the muon monitor is estimated at about 100~kGy for a 100-day operation with the \mbox{0.75-MW} proton beam power.

The muon intensity is monitored by measuring a sum of the 49 sensor signals.
The error on the intensity monitoring comes from the deviation from the detector's linearity against the muon intensity and stability, which have to be kept within 3\% in total.
The muon profile is measured by using the relative signal of each sensor.
Therefore, the error on the beam direction measurement comes from relative gain differences among the sensors in addition to the linearity and stability.
To achieve the requirement (2), each sensor should measure the muon intensity within 4\% precision.

\section{Instruments of the muon monitor} \label{sec:instruments}

\begin{figure}[t]
  \begin{center}
    \includegraphics[keepaspectratio=true,width=140mm]{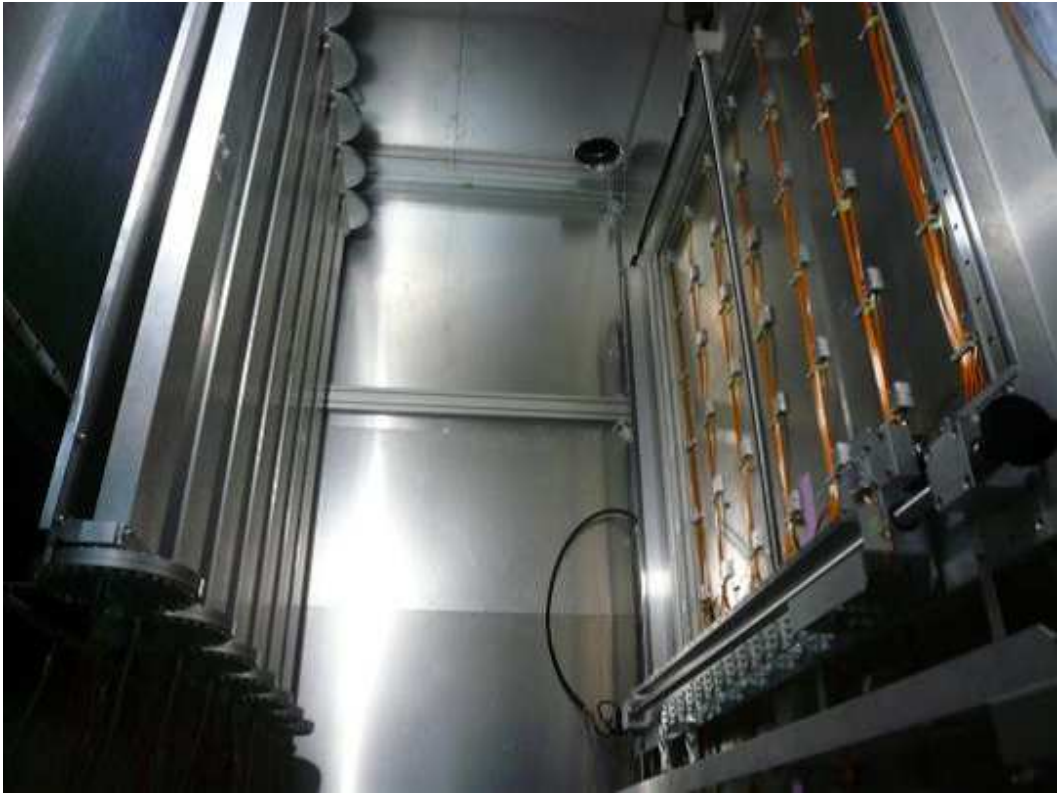}
  \end{center}
  \caption{Photograph of the silicon PIN photodiodes (right) and the ionization chambers (left) in the support enclosure.
  The beam enters from the right side.}
  \label{fig:photo MUMON}
\end{figure}

A schematic view of the muon monitor is shown in Fig.~\ref{fig:MUMON} and the photograph is in Fig.~\ref{fig:photo MUMON}.
The muon monitor consists of two independent detector systems: an array of ionization chambers and another array of silicon PIN photodiodes.
In each array, there are $7\times7$ sensors at \mbox{25-cm} intervals, also on the aluminum support enclosure.
The two different types of detectors provide redundant and complementary measurements.

Almost all components of the muon monitor in the muon pit are made out of radiation tolerant and low-activation materials such as polyimide, PEEK$^{\tiny \texttrademark}$, ceramic, aluminum and so on.
EPDM has the lowest radiation dose limit of $10^6$~Gy~\cite{dose limit} among the materials in the muon pit except the silicon PIN photodiodes.
The radiation tolerance of the photodiode is discussed in Sec.~\ref{sec:silicon} and \ref{sec:silicon radiation}.
Other instruments, like the readout electronics and high voltage (HV) units, are put in an electronics hut on the ground level.

\subsection{Ionization chamber}

\begin{figure}[t]
  \begin{center}
    \includegraphics[keepaspectratio=true,width=140mm]{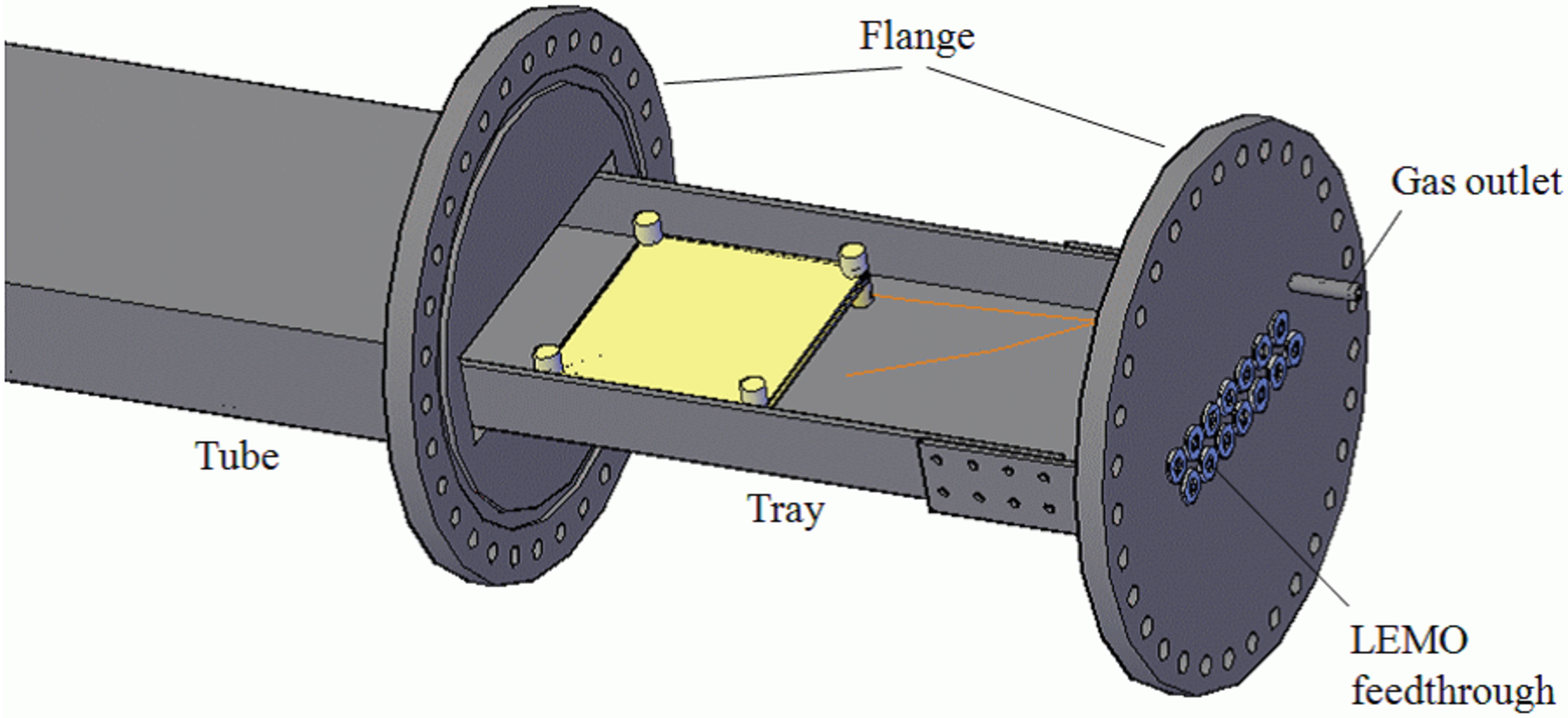}
  \end{center}
  \caption{Drawing of the bottom end of the ionization chamber.
  In this figure, the tray is pulled out of the tube and one of the seven sensors appears.}
  \label{fig:chamber}

  \begin{center}
    \includegraphics[keepaspectratio=true,width=140mm]{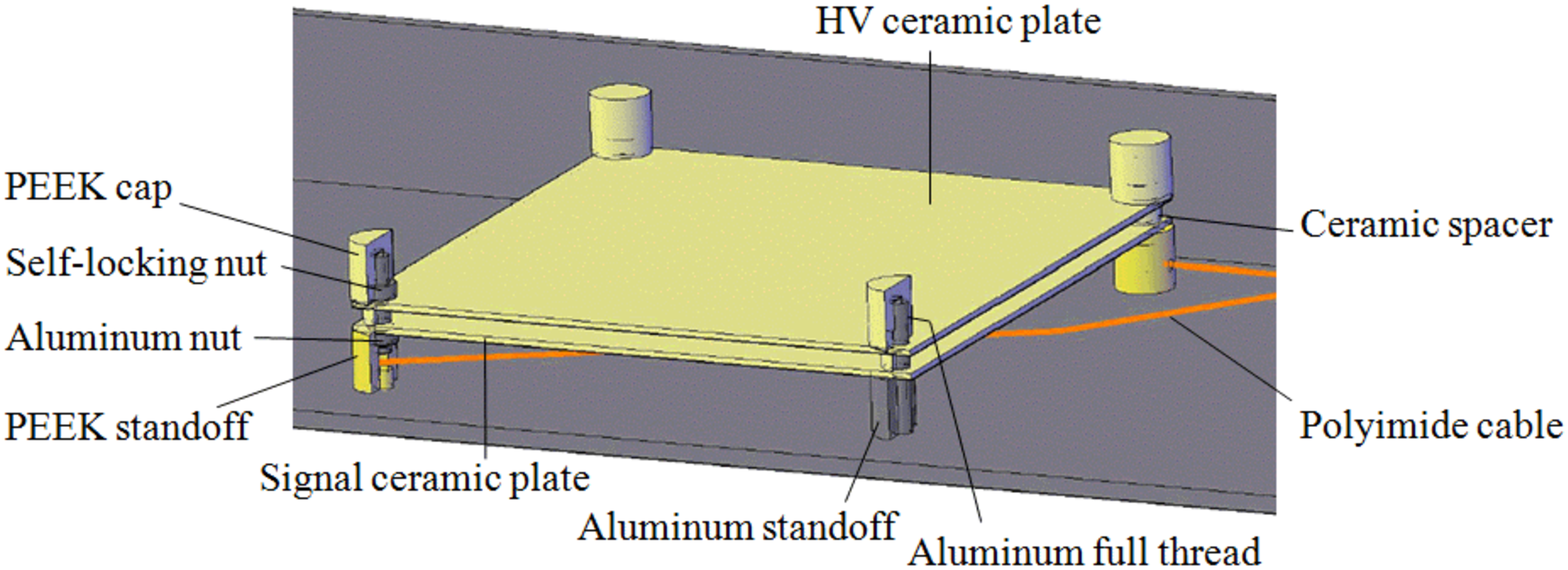}
  \end{center}
  \caption{Drawing of a sensor of the ionization chamber.
  The two sets of standoffs, ceramic spacers and PEEK caps are shown in section.}
  \label{fig:chamber plate}
\end{figure}

\subsubsection{Design of the ionization chamber} \label{sec:chamber design}
The ionization chamber array consists of seven ionization chambers.
A drawing of the ionization chamber is shown in Fig.~\ref{fig:chamber}.
Each ionization chamber contains seven sensors.
The sensors sit on a \mbox{1929-mm} long aluminum tray at an interval of 250~mm.
The tray is inserted into a $150\times50\times1956$~mm$^3$ aluminum tube.
At each end of the tube, aluminum flanges 235~mm in diameter are welded.
Other aluminum flanges are attached to them and seal the tube with O-rings (\mbox{U-TIGHT} SEAL$^{\tiny \textregistered}$ JIS \mbox{G-180}) made of Inconel$^{\tiny \textregistered}$ and aluminum.
Gas flows through \mbox{1/4-inch} tubes welded to the flanges.
On the bottom side flange, fourteen feedthrough connectors (LEMO$^{\tiny \textregistered}$ SWH.0S) for both signal and HV cables are mounted.
The LEMO feedthrough is sealed with an O-ring made of EPDM.
At the top in three out of the seven ionization chambers, a Pt100 resistance thermometer (PRT) in the four-wire configuration is installed to measure temperatures in the tubes.
For the PRTs, four-twisted polyimide-insulated wires and \mbox{D-sub} feedthrough connectors are used.
The insulators of those LEMO and \mbox{D-sub} feedthrough connectors are made of PEEK.

Each sensor of the ionization chamber consists of two parallel $100\times100\times1$~mm$^3$ alumina-ceramic plates, which are separated by 3~mm by ceramic spacers with a tolerance of 100~$\mu$m.
A drawing of the sensor is shown in Fig.~\ref{fig:chamber plate}.
Signal ($75\times75$~mm$^2$) and HV ($93\times93$~mm$^2$) electrodes made of \mbox{Ag-Pt} are printed on the ceramic plates.
The tolerance of the electrode size is 100~$\mu$m.
The signal electrode is surrounded by a grounded electrode, which ensures uniform electric field over the signal electrode.
Thus ionization pairs generated only in the $75\times 75\times 3$~mm$^3$ volume contribute to the signal.
Conductive parts except the \mbox{signal/HV} electrodes are insulated from the gas with the PEEK caps, ceramic spacers and PEEK standoffs in order to prevent stray ionization from collecting on the conductors~\cite{Bob Ph.D.}.

\subsubsection{Gas system}
Two kinds of gases are used for the ionization chambers: Ar with 2\% N$_2$ for a low intensity beam and He with 1\% N$_2$ for a high intensity beam.
By switching the gas from Ar to He, the signal is reduced by one order of magnitude because the ionization yields for a minimum ionization particle (MIP) in Ar and He gases at the standard condition\footnote{The temperature is 20$^\circ$C and the pressure is 101.325~kPa (absolute).} (STP) are 95.6~cm$^{-1}$ and 7.80~cm$^{-1}$~\cite{ICRU31}, respectively.
As a result, depletion of the signal due to recombination of electrons and ions can be avoided and the linearity of the signal to the beam intensity is guaranteed.
To get a faster response, only the charge induced by electrons' drift is used as a signal, and one by ions' drift ($10^3$ times slower) is disregarded.
In addition, N$_2$ is added as a quencher, which makes the response faster.
The N$_2$ plays another important role by making the Jesse effect~\cite{Jesse effect} saturated\footnote{If there is only a few impurities, there is a big increase in ionization due to Penning ionization.
Therefore, the signal increases as a function of the density of impurities.}, where the signal is insensitive to the amount of impurities in the gas.

\begin{figure}[bt]
  \begin{center}
    \includegraphics[keepaspectratio=true,width=140mm]{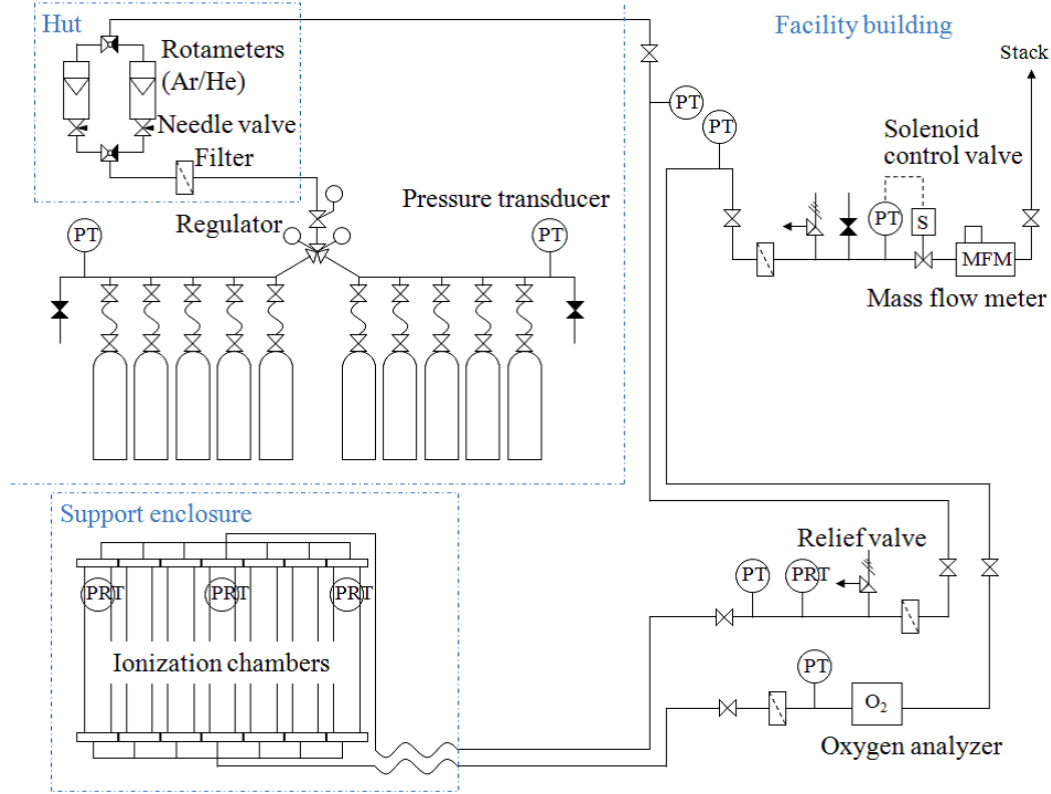}
  \end{center}
  \caption{Diagram of the gas system for the ionization chambers.
  PT stands for the pressure transducer and PRT stands for the Pt100 resistance thermometer.
  The dash-dotted lines shows the boundaries of the electronics hut, the facility building, and the support enclosure.}
  \label{fig:gas system}
\end{figure}

A diagram of the gas system is drawn in Fig.~\ref{fig:gas system}.
There are two feeds of the gas for the ionization chambers.
Each feed consists of a manifold of five \mbox{7-m$^3$} gas cylinders.
The gas pressure in each manifold is monitored by a pressure transducer (IBS$^{\tiny \textregistered}$ HSV-020MP).
At the outlet of the manifold, the gas is decompressed to 0.2~MPa by a regulator (Swagelok$^{\tiny \textregistered}$ KCM).
The regulator has a mechanism which automatically switches the gas feed from a depleted supply to the other and that ensures continuous flow of the gas.
Downstream of the regulator, the flow rate is set at approximately 100~cc/min.
Two rotameters for Ar and He gas are built into the line and they are exchangeable with ball valves at their inlets and outlets.
Proportional relief valves (Swagelok SS-RL3S4) in the line release the gas to protect the gas system if the pressure accidentally reaches 200~kPa (absolute).
Tee-type filters (Swagelok TF) protect the gas system from particulate contaminants.
Stainless tubes (1/4~inch) are used for the gas line to minimize outgassing from the tubes.

The stable response of the ionization chamber is guaranteed by a constant gas temperature, pressure and purity.
Variance of the temperature and pressure leads to variance in gas density, which varies the signal size.
Each parameter should be kept within 1.7\% to keep the signal variance within 3\%.
The gas temperatures in the ionization chambers are monitored by the three PRTs.
Temperatures in all of the ionization chambers are kept within 1.5$^\circ$C gradient and $\pm0.2^\circ$C variance at around 34$^\circ$C.
The absolute gas pressure is monitored by five pressure transducers (IBS HBV-300KP), which are kept separate from the muon pit to avoid radiation.
Two of them monitor the pressures in the feed and exhaust lines in the facility building, and two others monitor ones near the muon pit.
The fifth pressure transducer at the most downstream point is used for the PID control of a solenoid control valve (burkert$^{\tiny \textregistered}$ Type2822), which keeps the gas pressure in the entire gas line at $130\pm0.2$~kPa (absolute).
The O$_2$ contamination is the most important index of the gas purity since oxygen captures carrier electrons and decreases the signal.
It should be kept below 100~ppm in order to keep the signal depression by the electron attachment less than 1\%~\cite{electron attachment}.
The O$_2$ contamination is monitored by an oxygen analyzer (TORAY$^{\tiny \textregistered}$ LC-750L) in the exhaust line.
It is kept below 2~ppm at a gas flow rate of 100~cc/min.
The flow rate is monitored by a mass flow meter (HORIBASTEC$^{\tiny \textregistered}$ SEF-E40) just downstream of the control valve.

\subsection{Silicon PIN photodiode} \label{sec:silicon}
The silicon PIN photodiode (HAMAMATSU$^{\tiny \textregistered}$ S3590-08) has an active area of $10\times10$~mm$^2$ and a depletion layer thickness of 300~$\mu$m.
The silicon layer is mounted on a ceramic base.
To fully deplete the layer, a bias voltage of 80~V is applied.

The silicon PIN photodiode is not tolerant of the severe radiation in the muon pit:
The depletion voltage or the collected charge varies as a function of fluence.
There is a report~\cite{silicon radiation} that as the \mbox{647-MeV} proton fluence increases, the depletion voltage of their \mbox{200-$\mu$m} thick silicon PIN detectors decreases.
The depletion voltage falls 50\% at about $0.7\times 10^{13}$~protons/cm$^2$ and reaches a minimum at $1.25\times 10^{13}$~protons/cm$^2$, where type inversion occurs.
A \mbox{647-MeV} proton causes approximately the same amount of radiation damage as the \mbox{1-MeV} equivalent neutron fluence~\cite{1MeV neutron}.
For the T2K muon beam, the fluence is estimated at about $10^7$~\mbox{1-MeV}~neutrons/cm$^2$/spill at the beam center at 0.75~MW.
Therefore, the depletion voltage falls 50\% in about $0.7\times 10^{6}$~spills or a month.
Though the photodiode is not tolerant of radiation, it is useful at the early stage of the T2K experiment when the beam intensity is low and the noise/signal ratio of the ionization chamber is large.

Packages of the photodiodes were designed so that installation or replacement of the photodiodes can be quickly done in situ.
The photodiode is put on a PEEK base fixed on the support enclosure and is covered by an aluminum base.
In the PEEK base, two sockets soldered with co-axial polyimide cables are fixed.

\subsection{The support enclosure}\label{sec:support enclosure}
The dimensions of the support enclosure normal to and along the beam axis are 2.51~m and 1.71~m, respectively, and the height is 4.36~m.
The height of the beam axis at the silicon array is 2.87~m from the floor.
The secondary beamline was surveyed from the target station to the muon pit and the beam axis in the muon pit was defined with a precision of 1~cm.
The detectors are set with a precision better than 1~mm relative to the beam axis.

The support enclosure has two moving stages for relative calibration of the detectors: a small one which moves an extra silicon PIN photodiode just behind the silicon array and a large one which moves the ionization chambers.
The small moving stage is driven horizontally and vertically along linear guides by two stepping motors (ORIENTAL MOTOR RK569AMC).
On the linear guides, radiation-hard lubricant (MORESCO-HIRAD$^{\tiny \textregistered}$ RG-42-1) is used as grease.
The beam axis position, that is the origin of the stage, is detected by limit sensors of sliding touch type (Metrol$^{\tiny \textregistered}$ BP) assembled with polyimide-insulated wires.
The large stage for the ionization chambers moves horizontally and vertically along linear guides by two induction motors (ORIENTAL MOTOR BHI62SMT and 4IK25GN).
The position of the stage is detected by six limit sensors aligned horizontally and vertically at \mbox{25-cm} intervals.
Every limit sensor is aligned with a precision better than 1~mm.

The muon monitor is thermally insulated from the muon pit because the temperature in the muon pit could vary up to 33$^\circ$C\footnote{The variation of the temperature is due to a heat transfer from cooling water pipes for the beam dump.}.
Therefore, the support enclosure is covered by double aluminum panels filled with glass wool.
The temperatures around the ionization chambers are kept at around 34$^\circ$C by a sheathed heater and the variance of the temperatures is within $\pm0.7^\circ$C.

\subsection{Electronics and cables}
Signals of the detectors are read by \mbox{65-MHz} FADCs with a COPPER/FINESSE readout system~\cite{COPPER}.
The FADC gate opens for 7~$\mu$sec to contain all of the eight bunches.
The FADC has an input impedance of 50~$\Omega$ and an amplifier of \mbox{gain-5} or \mbox{gain-1} with a shaping time of 50~nsec.
The \mbox{gain-5} FADCs are used for the ionization chambers and the \mbox{gain-1} for the silicon PIN photodiodes.
The full scale of the FADC is $\pm1$~V after the amplifier and the dynamic range is 12~bits.
Because the signal of the silicon PIN photodiode is above the FADC full scale, it is attenuated with a $\pi$-type attenuator by 0, 15 or 30~dB according to the beam intensity.
The attenuator has linearity within $\pm0.9$\% up to 30~V of an input pulse and cross-talk of less than 0.02\% between adjacent channels.
A protection circuit is implemented for each channel and prevents surge voltage going to the FADC.
It consists of two zener diodes (RENESAS$^{\tiny \textregistered}$ HZ3A2) whose zener voltage is 2.6-2.8~V.
A series of the diodes in a reverse and forward bias order is connected parallel to the FADC input.
Thus a voltage loaded on the FADC is restricted within $\pm 4$~V.

HV units (REPIC$^{\tiny \textregistered}$ RPH-32010) are used to apply bias voltages to the detectors individually.
Low-pass filters follow the HV units to protect against surge voltage from the HV units and to ensure stable HV supply. 

In the muon pit, polyimide cables are used except for the power cables of the heater and the motors.
In the other area, non-halogen cables are used.
For example, co-axial non-halogen cables with 50-$\Omega$ impedance (\mbox{NH-5D-FB-E} for the signal and \mbox{RG-174/U$\times$10C} \mbox{NH-MCX} for the HV) are laid for about 60~m from the electronics hut to a patch panel at the entrance of the muon pit.
From the patch panel to the detectors, co-axial polyimide cables with 50-$\Omega$ impedance are laid for about 10~m both for the signal and the HV.

The FADCs and the signal cables were calibrated with a CAMAC charge/time generator (Phillips 7120) with a 1\% precision.

\section{Performance of the detectors}\label{sec:beam test}

Before installing the detectors in the T2K beamline, the ionization chambers and the silicon PIN photodiodes were tested with the \mbox{100-MeV} electron beam at the Institute for Chemical Research in Kyoto University to measure their charge collection times, dependence of the collected charge on the applied voltage and linearity of the collected charge to the beam intensity.
The detectors were also tested in the NuMI beamline to check the long-term stability of their response (FNAL T968 test experiment).
In addition, the radiation tolerance of the photodiode was examined by measuring the collected charge as a function of the electron beam fluence.

\subsection{Estimation of the detectors' signals} \label{sec:signal}

\begin{table}[b]
  \begin{center}
    \caption{Estimated collected charge $Q$ for the T2K muon beam at the intensity of $1\times10^7$~/cm$^2$/bunch with parameters in Eq.~\ref{eq:collected charge} ($\Delta_{ave}$, $\Delta_{mip}$ and $Y_{mip}$ for pure He and Ar gases; $Y'_{mip}$ and $Q$ for the He-N$_2$ and Ar-N$_2$ mixtures; $Y_{mip}$ and $Y'_{mip}$ at 34$^\circ$C and 130~kPa [absolute], and at STP in parentheses).}
    \label{tbl:signal}
    \vspace{\baselineskip}
    \begin{tabular}{|l|c|c|c|}
      \hline
       & Helium & Argon & Silicon \\
      \hline
      $\Delta_{ave}$ (MeV~g$^{-1}$cm$^2$)~\cite{PDG} & 2.427 & 1.980 & 1.991 \\
      \hline
      $\Delta_{mip}$ (MeV~g$^{-1}$cm$^2$)~\cite{PDG} & 1.937 & 1.519 & 1.664 \\
      \hline
      $Y_{mip}$ (pairs/cm) & 9.55 (7.80)~\cite{ICRU31} & 117 (95.6)~\cite{ICRU31} & $9\times 10^5$ \\
      \hline
      $Y'_{mip}$ (pairs/cm) & 14.3 (with N$_2$) & 117 (with N$_2$) &  \\
      \hline
      $Q$ (nC/bunch) & 0.242 (with N$_2$) & 2.06 (with N$_2$) & 52 \\
      \hline
    \end{tabular}
  \end{center}
\end{table}

The detector's collected charge $Q$ can be estimated by using the following equation:
\begin{equation} \label{eq:collected charge}
  Q = \left\{
  \begin{array}{ll}
    N e_0 (\Delta_{ave} / \Delta_{mip}) Y_{mip} V /2 \; & \mbox{for the ionization chamber,} \\
    N e_0 (\Delta_{ave} / \Delta_{mip}) Y_{mip} V       & \mbox{for the silicon PIN photodiode,}
  \end{array}
  \right.
\end{equation}
where $N$ is the number of charged particles, $e_0$ is the electron charge, $\Delta_{ave}$ and $\Delta_{mip}$ is the ionization energy losses by a muon of average energy in the T2K muon beam and by a MIP, respectively, $Y_{mip}$ is the ionization yield for a MIP, and $V$ is the active volume of the detectors.
For the ionization chamber, the right side of Eq.~\ref{eq:collected charge} is divided by two since only the charge induced by electrons is used.
The average energy of the muons at the muon pit is estimated at approximately 3~GeV.

The values in Eq.~\ref{eq:collected charge} for He, Ar and silicon are listed in Table~\ref{tbl:signal}.
$Y_{mip}$ for the He-N$_2$ mixture (expressed as $Y'_{mip}$) is larger than one for pure He gas by about 50\%~\cite{Jesse effect} due to the Jesse effect, while $Y_{mip}$ for the Ar-N$_2$ mixture is almost same as one for Ar gas.
$Q$ of the He-N$_2$ and Ar-N$_2$ ionization chambers and the silicon PIN photodiode are estimated at 0.242, 2.06 and 52~nC/bunch, respectively, for the muon beam intensity of $1\times10^7$~/cm$^2$/bunch.

The collected charges for the electron beam are not the same as ones for the T2K muon beam discussed above.
In Eq.~\ref{eq:collected charge}, $\Delta_{ave}$ can be replaced with the ionization energy loss $\Delta_{e}$ by a \mbox{100-MeV} electron in He, Ar and silicon: 2.609, 2.105 and 1.919~MeV~g$^{-1}$cm$^2$~\cite{ESTAR}, respectively.
Therefore, the difference of the collected charge is given as $\Delta_{e} / \Delta_{ave}$: 1.075 for He, 1.063 for Ar and 0.964 for silicon.

\subsection{Setup of the beam tests}

\subsubsection{Test configuration at the electron beamline}
The characteristics of the electron beam are listed in Table~\ref{tbl:beam spec}.
The linear accelerator produced the \mbox{100-MeV} electron beam with a repetition rate of 15~Hz.
The beam intensity was $10^7$~to~$10^9$~electrons/pulse and the width was 60~nsec.
They were adjusted to those of the expected T2K muon beam.
Because the profile of the electron beam was close to a Gaussian distribution with a sigma of 0.6~cm,
the signal electrode of the ionization chamber received the whole beam.
The fraction of the electrons passing through the silicon PIN photodiode was estimated from the profile.

Detectors were placed in the beamline in the following order from the upstream: current transformers (CTs) for intensity monitors,
an array of nine silicon PIN photodiodes for the profile monitor and a sensor of the ionization chamber.
Both the Ar with 2\% N$_2$ and He with 1\% N$_2$ were tested at room temperature, atmospheric pressure and a flow rate of approximately 100~cc/min.
Variation of the temperature and the pressure during the measurement was negligible.
The O$_2$ contamination in the gas was less than 100~ppm.

Signals of the detectors were read by the COPPER/FINESSE FADCs.
Signals of the photodiodes were attenuated by 16 to 46~dB with variable attenuators.

\begin{table}[b]
  \begin{center}
    \caption{Characteristics of the electron beam and the NuMI muon beam at muon alcove~2.
    Those for the expected T2K muon beam are also listed for comparison.
    The intensity of the electron beam is adjustable and the maximum used value averaged over the active area of the ionization chamber is listed.
    The intensity of the NuMI muon beam is the nominal value for the test and the T2K muon beam is an expectation of its maximum intensity.}
    \label{tbl:beam spec}
    \vspace{\baselineskip}
    \begin{tabular}{|l|c|c|c|}
      \hline
       & Electron & NuMI muon & T2K muon \\
      \hline
      Intensity [$10^7$/cm$^2$/bunch(batch)] &  $< 2.3$ & 0.03 & 1 \\
      \hline
      No. of bunches (batches) in a spill & 1 & 5 or 6 & 8 \\
      \hline
      Bunch (batch) width [sec] & 60 n & 1.6 $\mu$ & 58 n \\
      \hline
      Spill repetition [Hz] & 15 & 0.53 & 0.285 \\
      \hline
      Beam size ($\sigma$) [m] & 0.006 & 0.8 & 1 \\
      \hline
    \end{tabular}
  \end{center}
\end{table}

\subsubsection{Test configuration at the NuMI beamline (FNAL T968)}
The NuMI beamline is operated for the MINOS (Main Injector Neutrino Oscillation Search) experiment~\cite{MINOS}.
The typical beam intensity during our measurement was $2\times10^{13}$~protons-per-pulse (ppp) with a repetition rate of 0.53~Hz.
A spill contains 5 or 6 batches in 8.14 or 9.78~$\mu$sec, respectively, and a batch contains 84 bunches.
The width of the bunch is 3~to~8~nsec.
We installed a miniature ionization chamber and two silicon PIN photodiodes just behind the NuMI muon monitor~\cite{NuMI MUMON} in the muon alcove~2.
The characteristics of the beam there~\cite{NuMI MUMON} is listed in Table~\ref{tbl:beam spec}.
They estimated the intensity of charged particles there at $1.7\times10^6$~/cm$^2$/spill at the profile center for $2\times10^{13}$~ppp.

The miniature chamber has the same design as the ionization chamber of the T2K muon monitor, but only three sensors.
It was hung vertically and the top sensor was near the center of the muon beam.
The exhaust pure He gas from the NuMI muon monitor 2 was supplied to the miniature chamber.
The gas pressure, temperature and flow rate were at atmospheric pressure, room temperature and approximately 500~cc/min, respectively.
The O$_2$ contamination at the outlet of the miniature chamber was measured once after the installation and was 52.0~ppm.
The silicon PIN photodiodes were put on the rear of the miniature chamber.
Their centers were aligned with those of the top two sensors of the miniature chamber.

Signals were read by a CAMAC charge ADC (LeCroy$^{\tiny \textregistered}$ 2249W).
Signals of the photodiodes were attenuated by 20~dB with co-axial type attenuators.

\subsection{Results of the beam tests}

\subsubsection{Charge collection time} \label{sec:Qt}

\begin{figure}[tb]
  \begin{minipage}{0.5\hsize}
    \begin{center}
      \includegraphics[keepaspectratio=true,width=70mm]{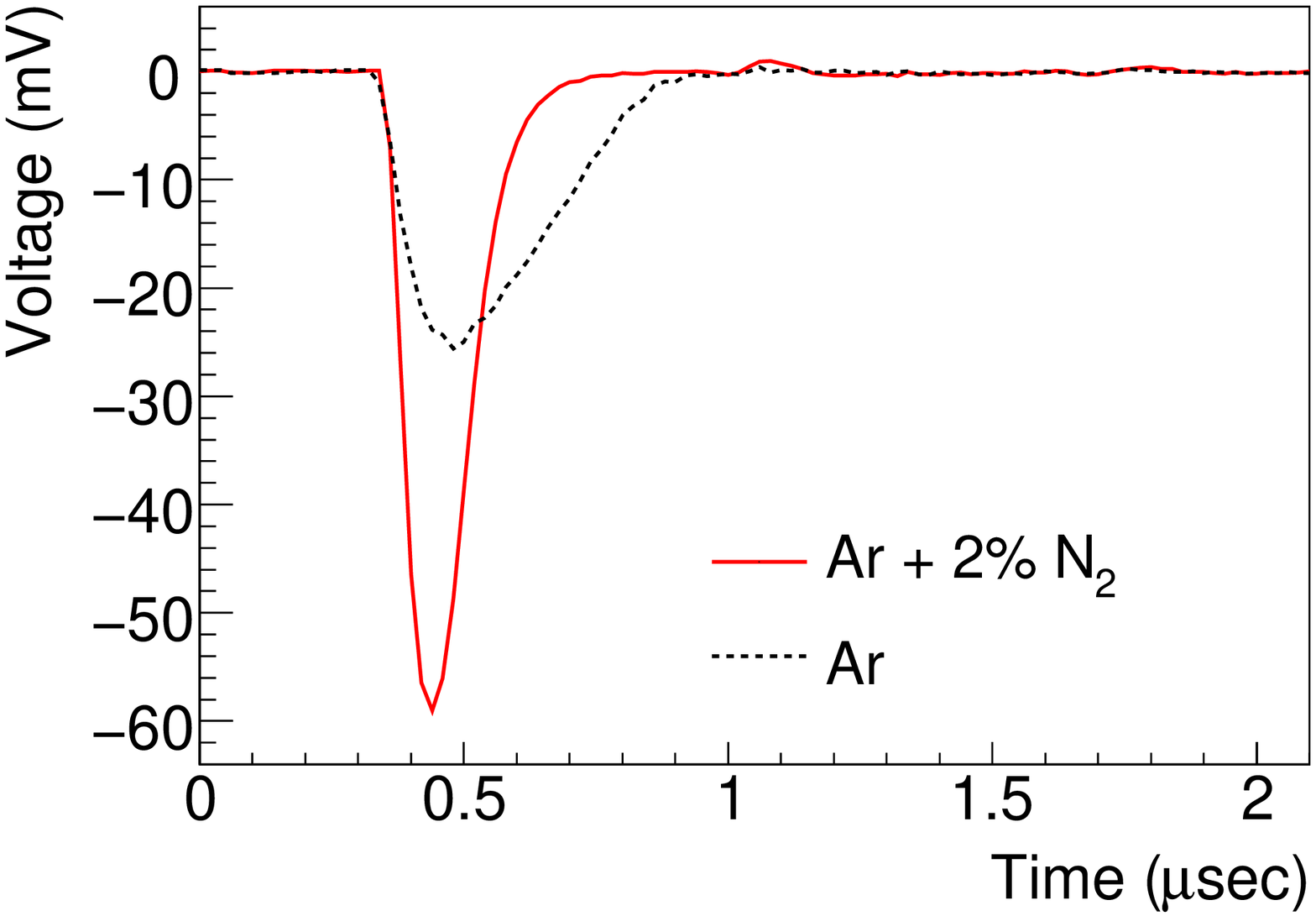}
    \end{center}
  \end{minipage}
  \begin{minipage}{0.5\hsize}
    \begin{center}
      \includegraphics[keepaspectratio=true,width=70mm]{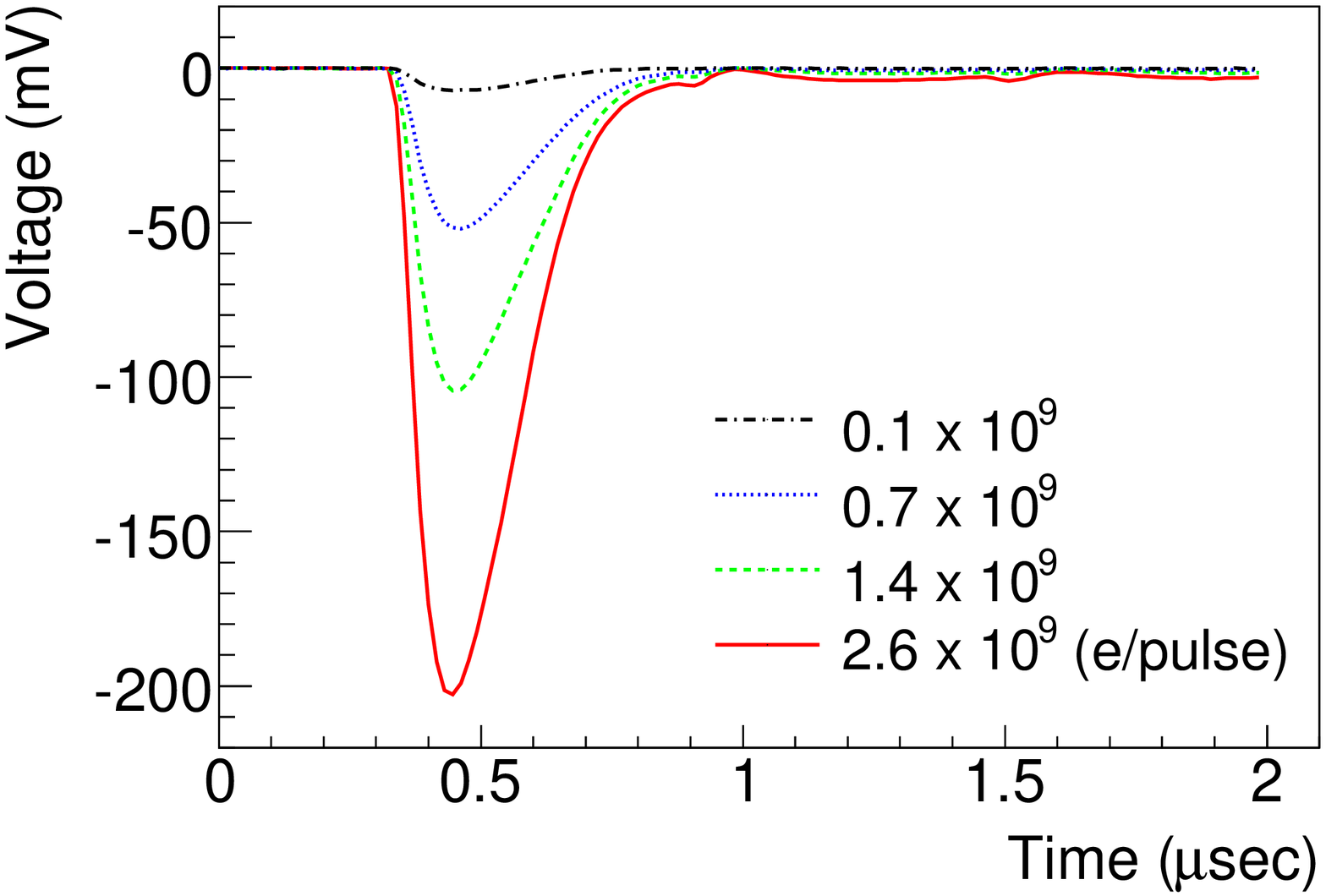}
    \end{center}
  \end{minipage}
  \caption{\label{fig:pulse shape IC}
  Left: pulse shapes of the ionization chamber with the Ar-N$_2$ mixture and with pure Ar gas read by the FADC with the applied voltage of 200~V and at the same beam intensity of $8\times10^7$~electrons/pulse.
  Right: pulse shapes of the ionization chamber with the He-N$_2$ mixture at 200~V and at several beam intensities.}
  \vspace{20pt}

  \begin{center}
    \includegraphics[keepaspectratio=true,width=70mm]{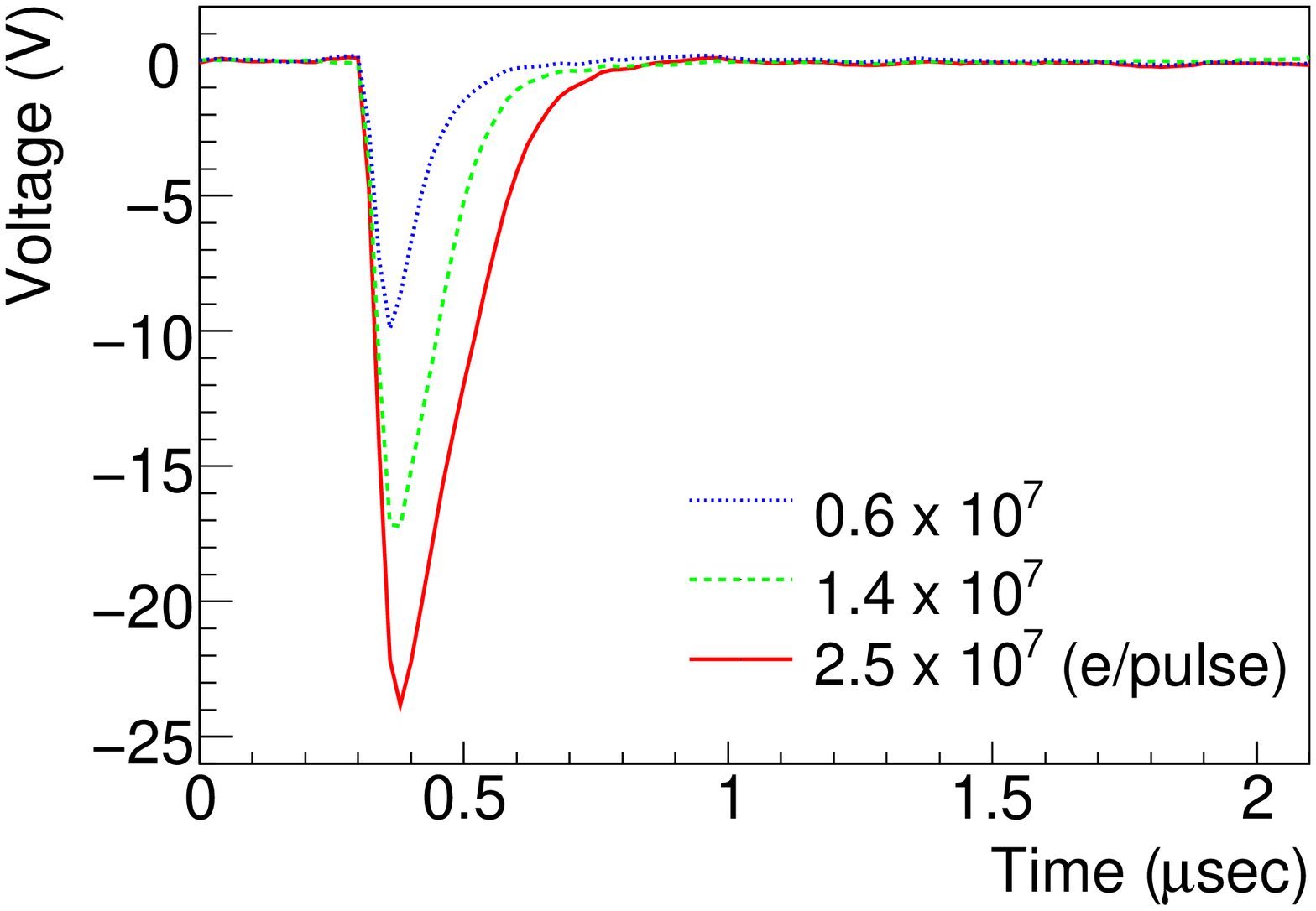}
  \end{center}
  \caption{\label{fig:pulse shape Si}
  Pulse shapes of the silicon PIN photodiode read by the FADC with the applied voltage of 80~V and at several beam intensities.
  Intensities passing through the full active area of the photodiode are written in the figure.}
\end{figure}

The full widths of the detectors' pulses were measured.
They have to be shorter than 581~nsec (the T2K bunch interval) for the bunch-by-bunch measurement of the T2K muon beam.

The left graph in Fig.~\ref{fig:pulse shape IC} shows pulse shapes of the ionization chamber with the Ar-N$_2$ mixture and with pure Ar gas with the applied voltage of 200~V and at the same beam intensity of $8\times10^7$~electrons/pulse.
Adding 2\% N$_2$ makes the response twice as fast as that of pure Ar.
The full pulse width of the Ar-N$_2$ chamber is 400~nsec.
The right graph in Fig.~\ref{fig:pulse shape IC} shows pulse shapes of the ionization chamber with the He-N$_2$ mixture at 200~V and at several beam intensities.
The full pulse width is 700~nsec, but the fraction of the pulse pile up on the following bunch is less than 1\%.

Figure~\ref{fig:pulse shape Si} shows pulse shapes of the silicon PIN photodiode with the applied voltage of 80~V and at several beam intensities.
Both the pulse height and width depend on the beam intensity.
That is because the drift velocity decreases with the depletion of the electric potential between the electrodes when there is a large charge induced by the drift of the ionization electron.
The full pulse width is 500~nsec at $1.4\times10^7$~electrons/pulse.

\subsubsection{Dependence of the collected charge on the applied voltage} \label{sec:Q-V}
\begin{figure}[tb]
  \begin{minipage}{0.5\hsize}
    \begin{center}
      \includegraphics[keepaspectratio=true,width=70mm]{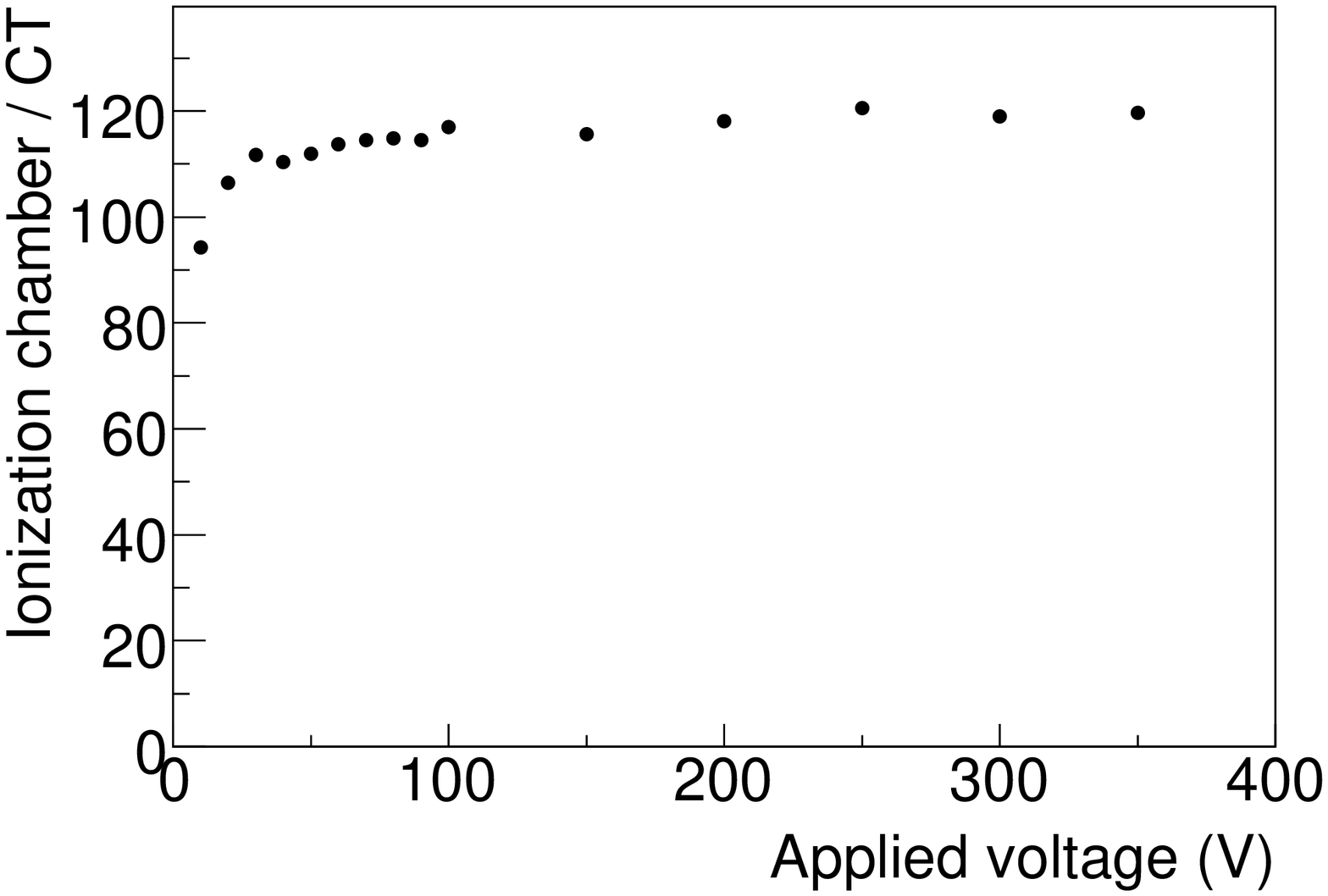}
    \end{center}
  \end{minipage}
  \begin{minipage}{0.5\hsize}
    \begin{center}
      \includegraphics[keepaspectratio=true,width=70mm]{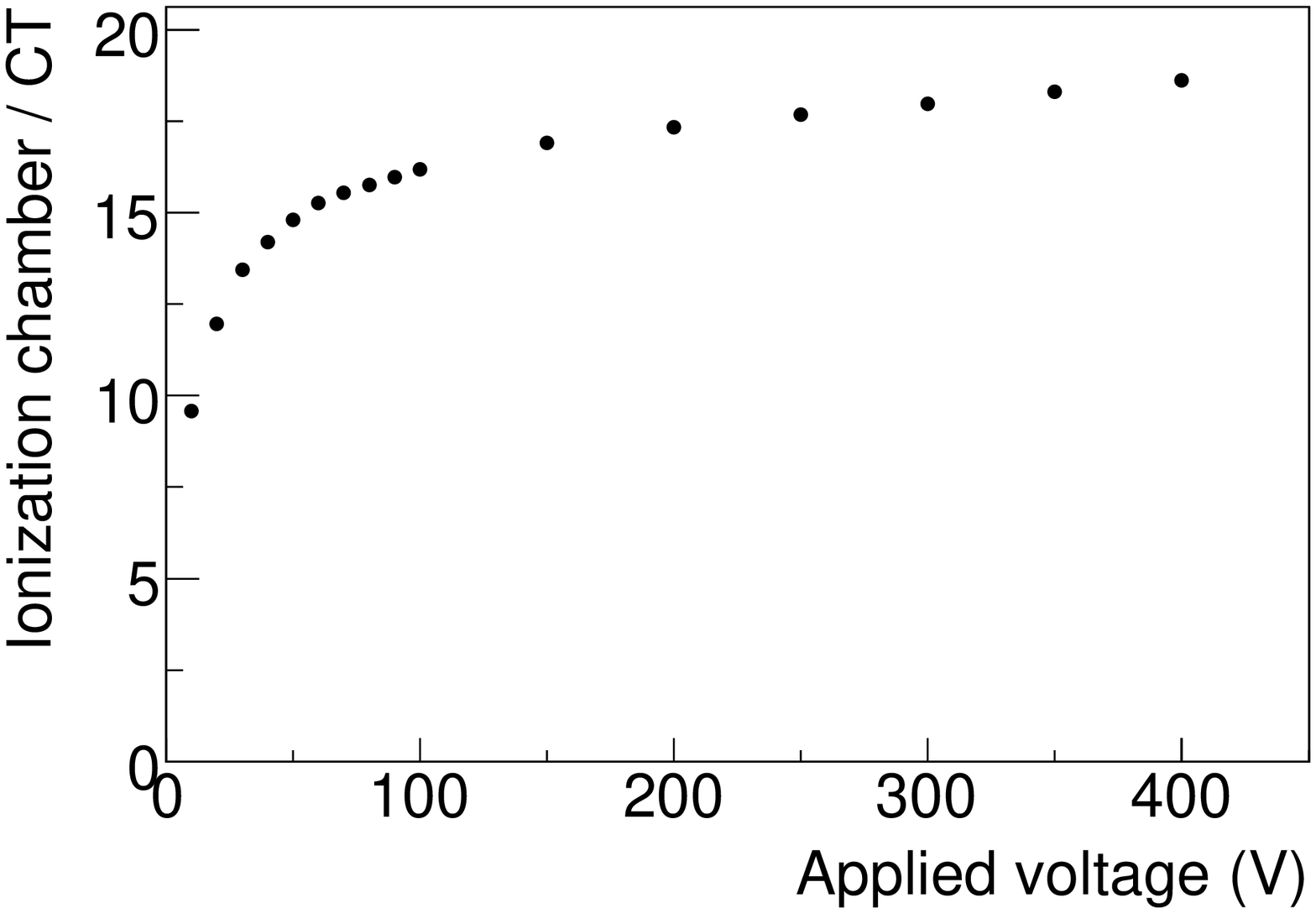}
    \end{center}
  \end{minipage}
  \caption{\label{fig:HV curve IC}
  Collected charge as a function of the applied voltage for the ionization chamber with the Ar-N$_2$ (left) and with the He-N$_2$ mixtures (right).
  The collected charge of the ionization chamber is normalized by the CT signal.}
\end{figure}

Figure~\ref{fig:HV curve IC} shows the collected charge of the ionization chamber normalized by the CT signal as a function of the applied voltage.
The beam intensity was kept at $3\times10^7$~electrons/pulse for the Ar-N$_2$ mixture and $6\times10^8$~electrons/pulse for the He-N$_2$ mixture.
The signal of the Ar-N$_2$ ionization chamber in the left graph gets closer to a constant value as the applied voltage increases. 
On the other hand, the signal of the He-N$_2$ ionization chamber in the right graph does not, although the increase is negligibly as small as 0.04\%/V at 200~V.
The increase is due to the contribution by ions\footnote{For a \mbox{3-mm} gap and at 200~V, drift times of He$_2^+$, He$^+$ ions and an electron in 0$^\circ$C \mbox{1-atm} He gas are 23~$\mu$sec~\cite{ion mobility}, 40~$\mu$sec~\cite{ion mobility} and 380~nsec~\cite{electron drift He}, respectively, while drift times of Ar$_2^+$, Ar$^+$ ions and an electron in Ar gas are 0.17~msec~\cite{ion mobility}, 0.27~msec~\cite{ion mobility} and 750~nsec~\cite{electron drift Ar}, respectively.};
about 10\% of ions from ionization pairs contribute to the signal at 200~V and the contribution increases as the applied voltage increases, while the electrons are fully collected.

\subsubsection{Linearity} \label{sec:linearity}
\begin{figure}[tb]
  \begin{minipage}{0.5\hsize}
    \begin{center}
      \includegraphics[keepaspectratio=true,width=70mm]{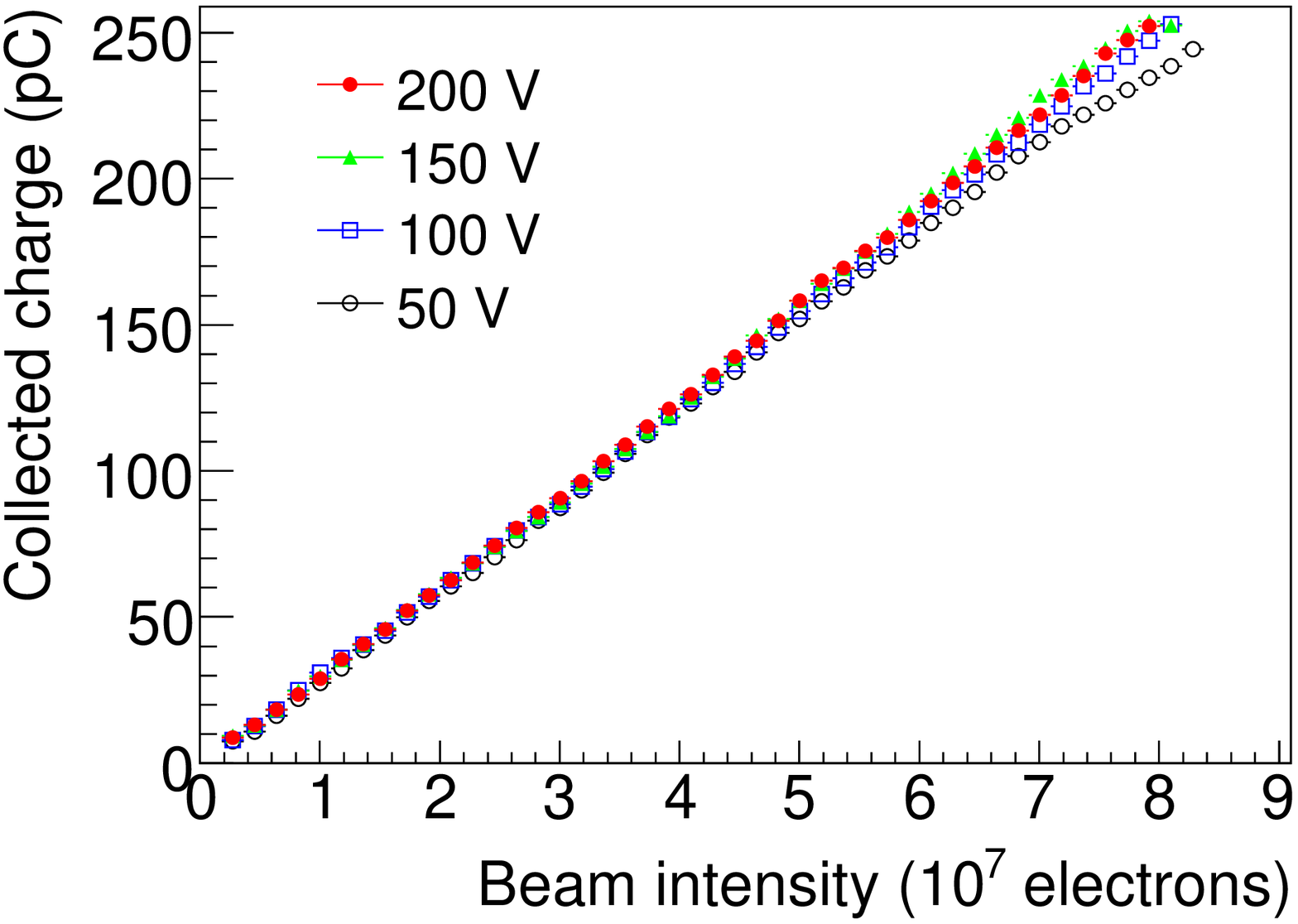}
    \end{center}
  \end{minipage}
  \begin{minipage}{0.5\hsize}
    \begin{center}
      \includegraphics[keepaspectratio=true,width=70mm]{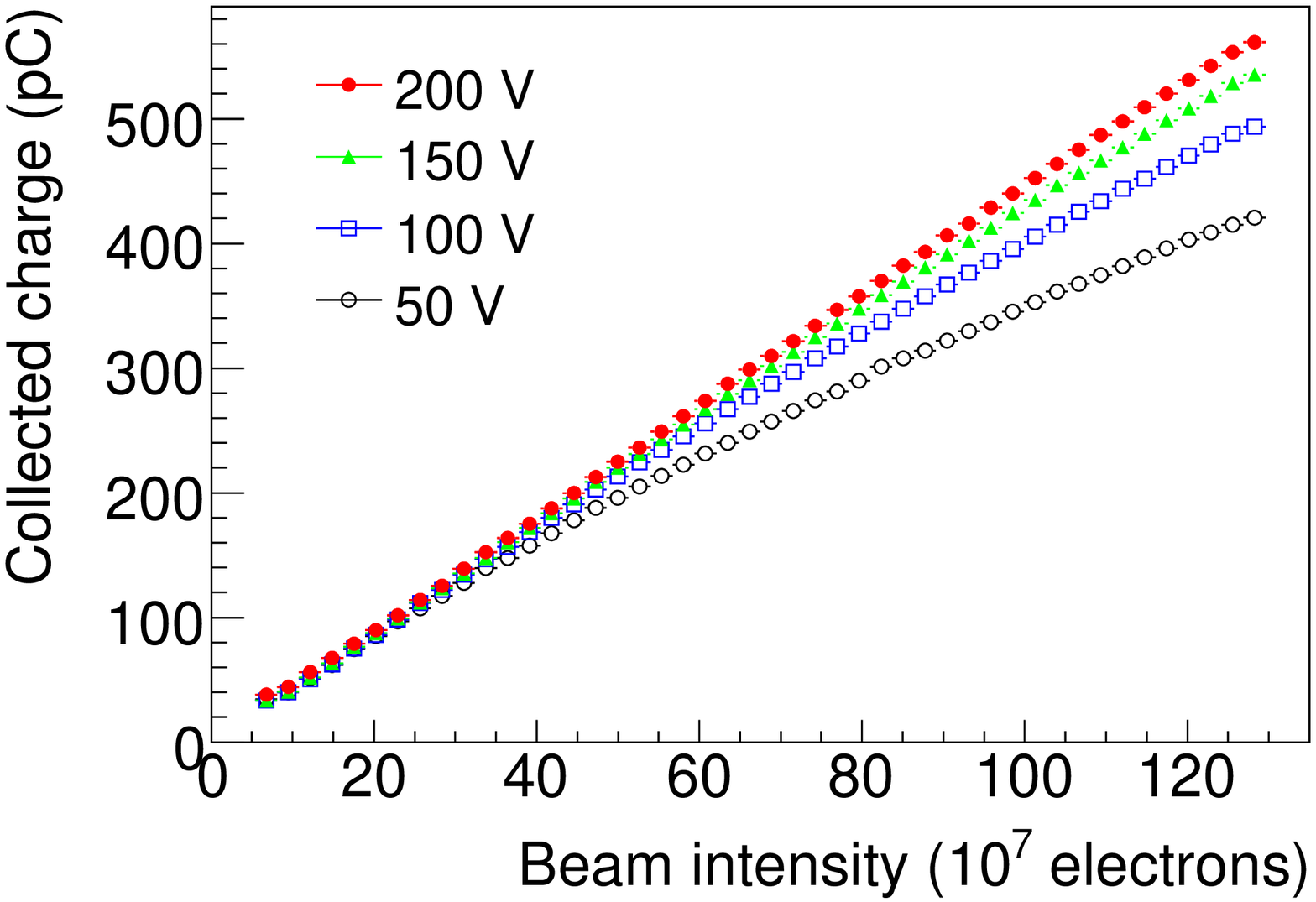}
    \end{center}
  \end{minipage}
  \caption{\label{fig:linearity IC}
  Collected charge to the beam intensity for the ionization chamber with the Ar-N$_2$ (left) and with He-N$_2$ mixtures (right) at several applied voltages.}
  \vspace{20pt}

  \begin{center}
    \includegraphics[keepaspectratio=true,width=70mm]{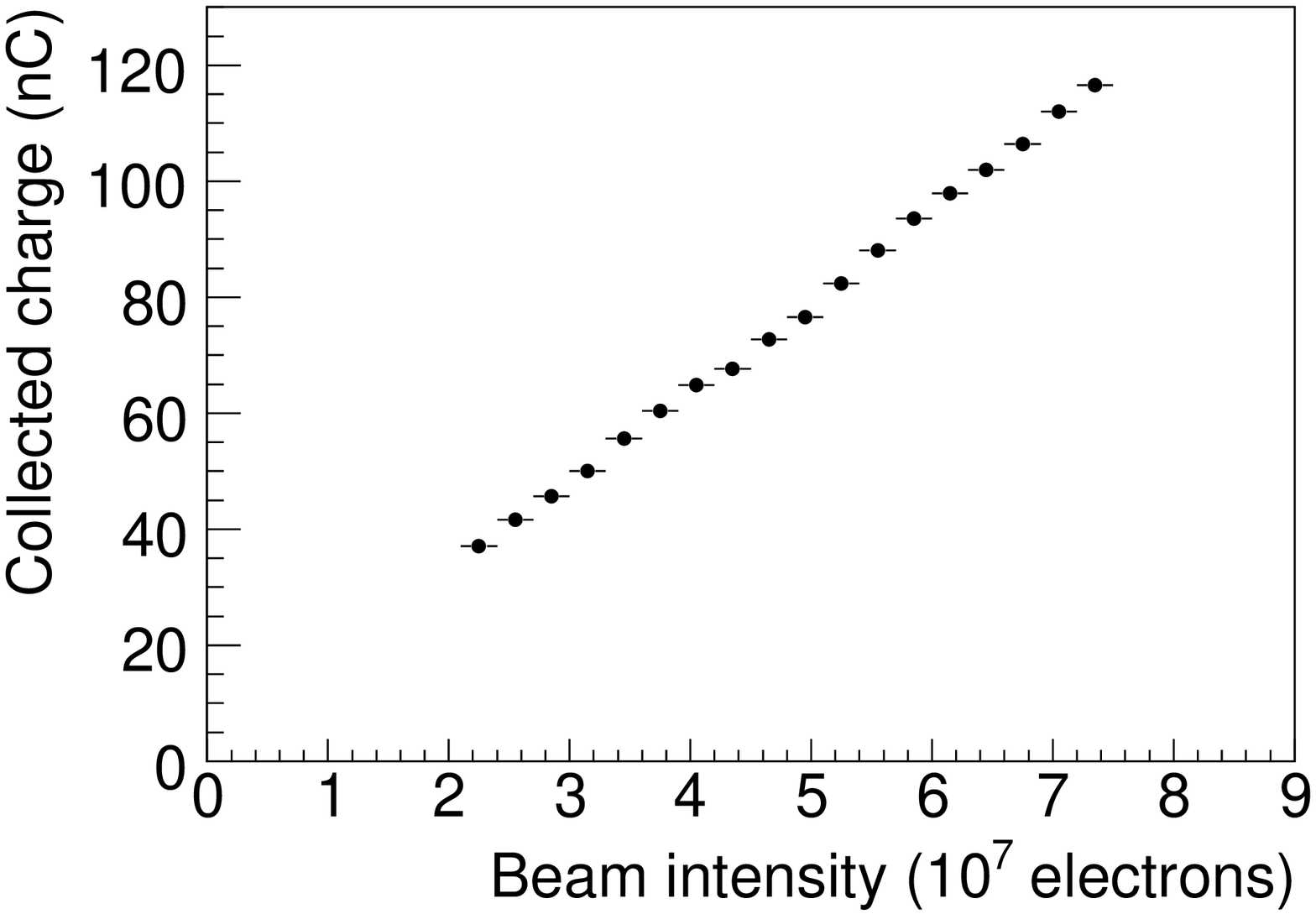}
  \end{center}
  \caption{\label{fig:linearity Si}
  Collected charge to the beam intensity for the silicon PIN photodiode.
  Approximately 30\% of the beam enters the active area.}
\end{figure}

Figure~\ref{fig:linearity IC} shows the collected charge of the ionization chamber versus the beam intensity with several applied voltages.
For the Ar-N$_2$ mixture, the beam intensity was measured by the nine silicon PIN photodiodes of the profile monitor instead of the CT since the CT signal was too small to be measured.
The response is linear to the beam intensity at voltages above 100~V for the Ar-N$_2$ and at 200~V for the He-N$_2$ ionization chamber.
For the He-N$_2$ ionization chamber at the lower applied voltages, recombination can be clearly seen at the high intensities.
The residual to a linear line fit is less than $\pm 2.4$\% up to $1.4\times 10^6$ electrons/cm$^2$\footnote{The average beam intensity over the signal electrode.} for the Ar-N$_2$ and $\pm 1.7$\% up to $2.3\times 10^7$ electrons/cm$^2$ for the He-N$_2$ ionization chamber at 200~V.

The electron beam concentrates around the center of the signal electrode of the ionization chamber, while the T2K muon beam uniformly passes over the electrode.
Because the probability of recombination is proportional to the density of ions and electrons, the number of recombinations is proportional to $\int_{S} f(x,y)^2 dxdy$, where $f(x,y)$ is the beam density and $S$ is the active area.
The number of recombinations is approximately 12~times larger for the electron beam than for the first bunch of the T2K muon beam.
In the case of the T2K multi-bunch beam, the number of recombinations increases in proportion to the number of bunches because ions produced by the preceding bunches stay between the electrodes due to their long drift time.
Therefore, the number of recombinations in this test is equivalent to one at the twelfth bunch.
Thus the linearity at 200~V is guaranteed for all eight bunches in T2K as a result of this test.

Figure~\ref{fig:linearity Si} shows the collected charge of the silicon PIN photodiode at the center of the profile monitor versus the beam intensity.
The fraction of the electrons passing through the full active area is estimated at approximately 30\%.
The response is linear to the beam intensity.
The residual to a linear line fit is less than $\pm 1.9$\% up to $2.2\times 10^7$ electrons/cm$^2$.

The collected charge of the ionization chamber at 200~V is approximately 250~pC at $7.9\times 10^7$~electrons/pulse for the Ar-N$_2$ mixture, while the rough estimation at STP with the correction of energy loss by the electron/muon beam was 250~pC as discussed in Sec.~\ref{sec:signal}.
For the He-N$_2$ mixture, the collected charge is approximately 250~pC at $5.6\times 10^8$~electrons/pulse and the estimated one was 234~pC (including 10\% of the ions' contribution).
For the silicon PIN photodiode, the collected charge is approximately 53~nC at $3.3\times 10^7$~electrons/pulse and the estimated one was 50~nC.
The measured and estimated charges agree well, which ensures that the detectors work as designed.

\subsubsection{Long-term stability of the response} \label{sec:stability}
\begin{figure}[bt]
  \begin{minipage}{0.5\hsize}
    \begin{center}
      \includegraphics[keepaspectratio=true,width=70mm]{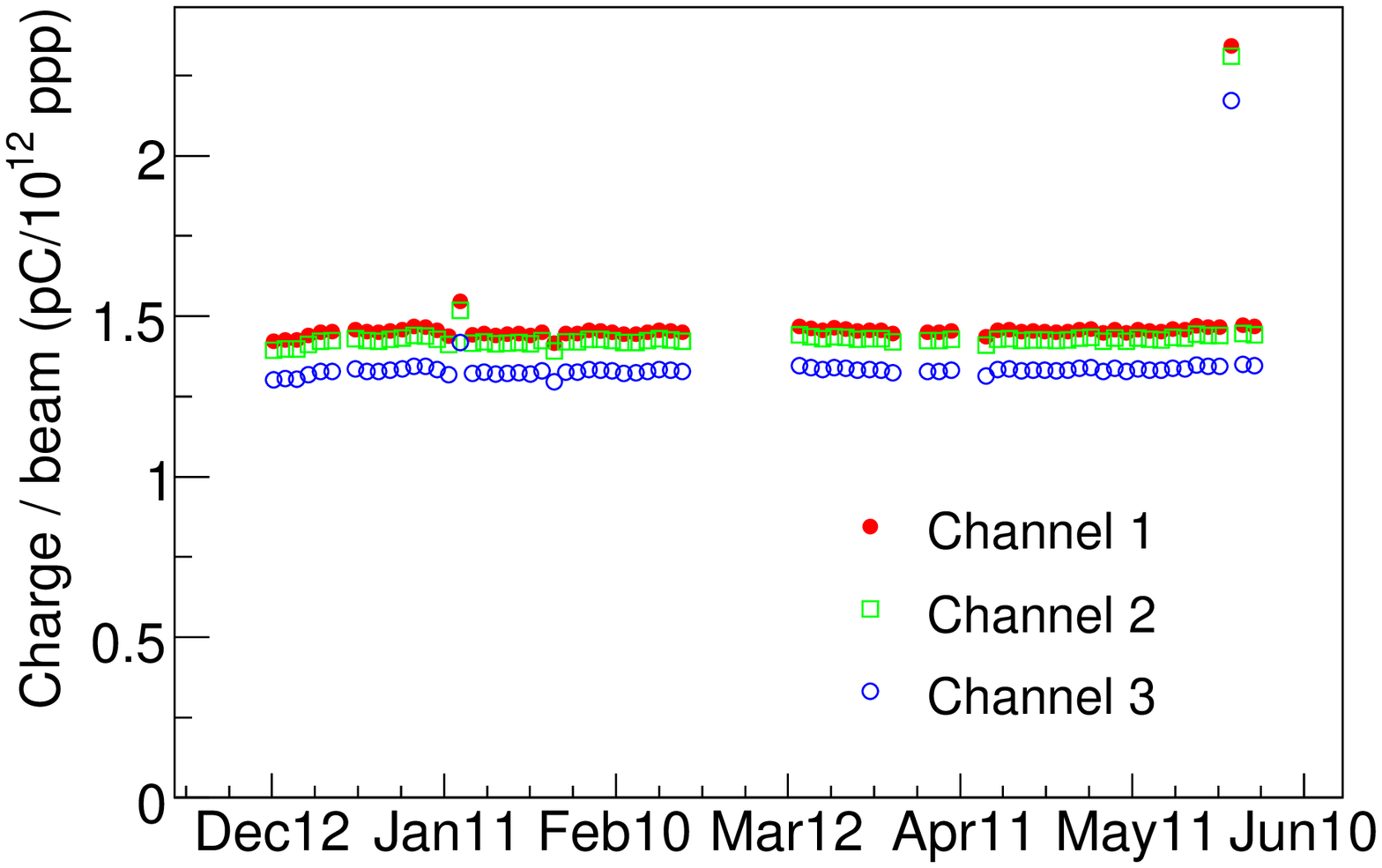}
    \end{center}
  \end{minipage}
  \begin{minipage}{0.5\hsize}
    \begin{center}
      \includegraphics[keepaspectratio=true,width=70mm]{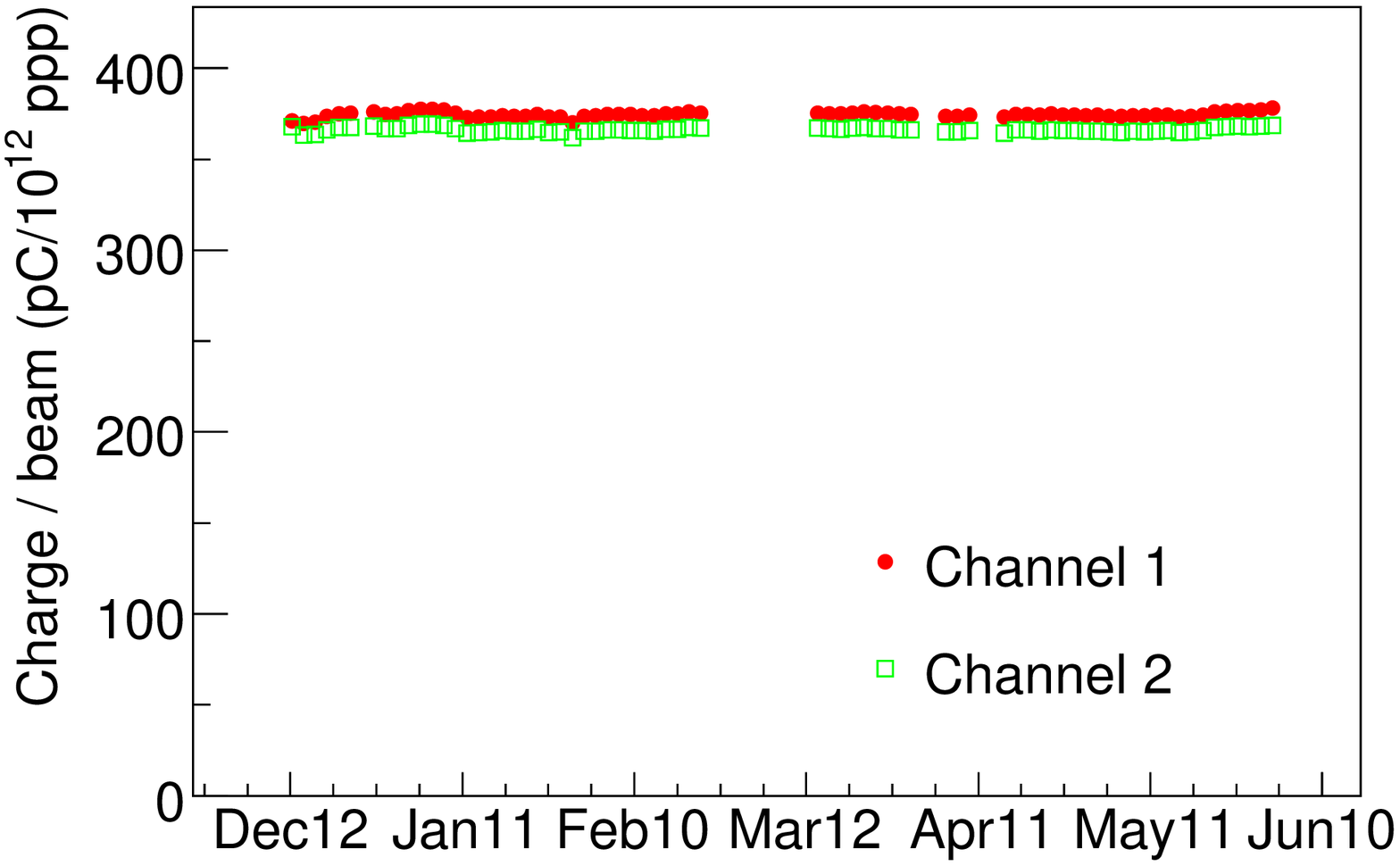}
    \end{center}
  \end{minipage}
  \caption{\label{fig:stability}
  Stability of the signal of the miniature ionization chamber with the pure He gas (left) and the silicon PIN photodiodes (right) measured in the NuMI beamline.
  The collected charge of the detectors is normalized by the proton intensity.
  The two spikes in the ionization chamber charges in January and May are due to deterioration of the gas purity.}
  \vspace{20pt}

  \begin{minipage}{0.5\hsize}
    \begin{center}
      \includegraphics[keepaspectratio=true,width=70mm]{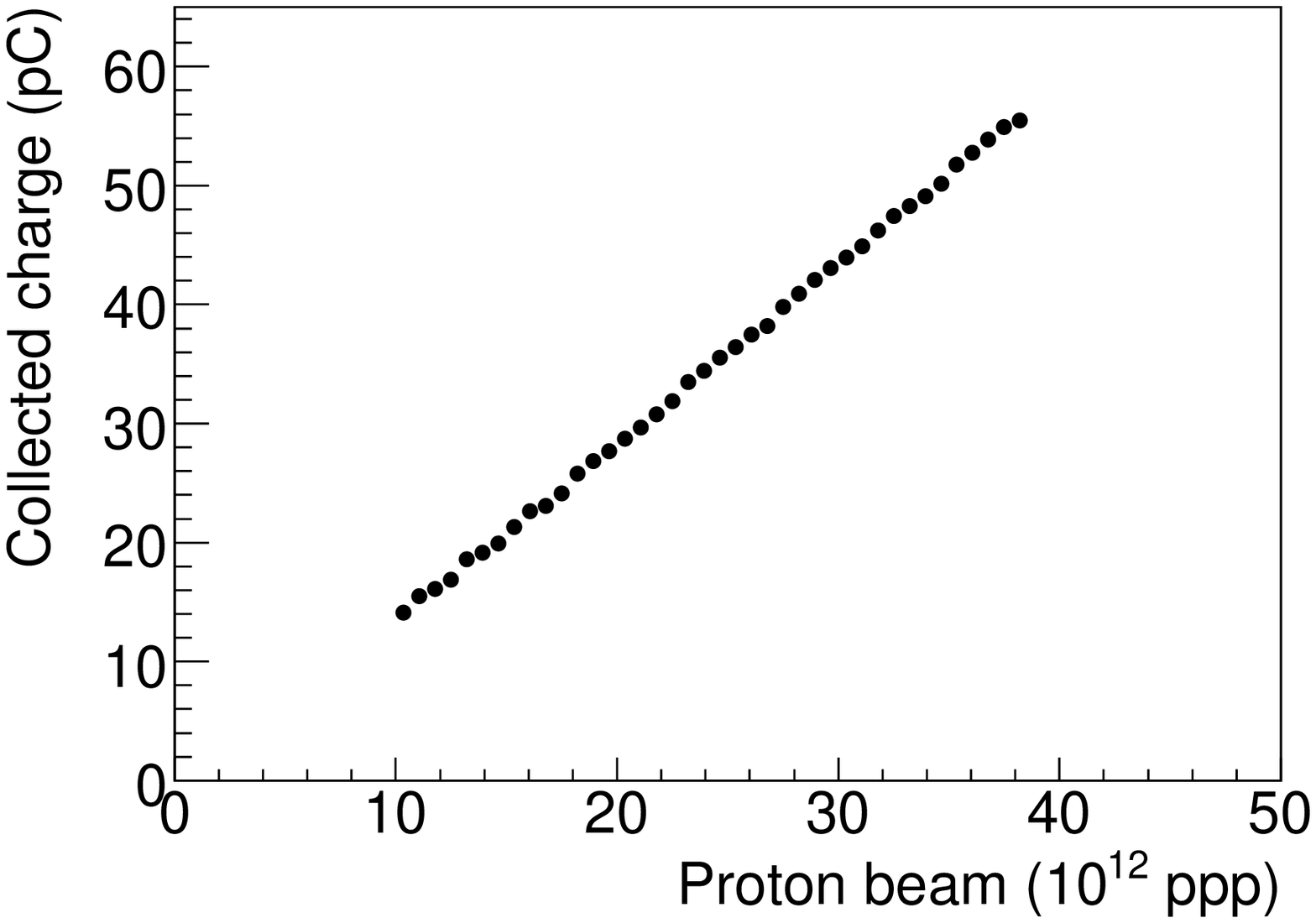}
    \end{center}
  \end{minipage}
  \begin{minipage}{0.5\hsize}
    \begin{center}
      \includegraphics[keepaspectratio=true,width=70mm]{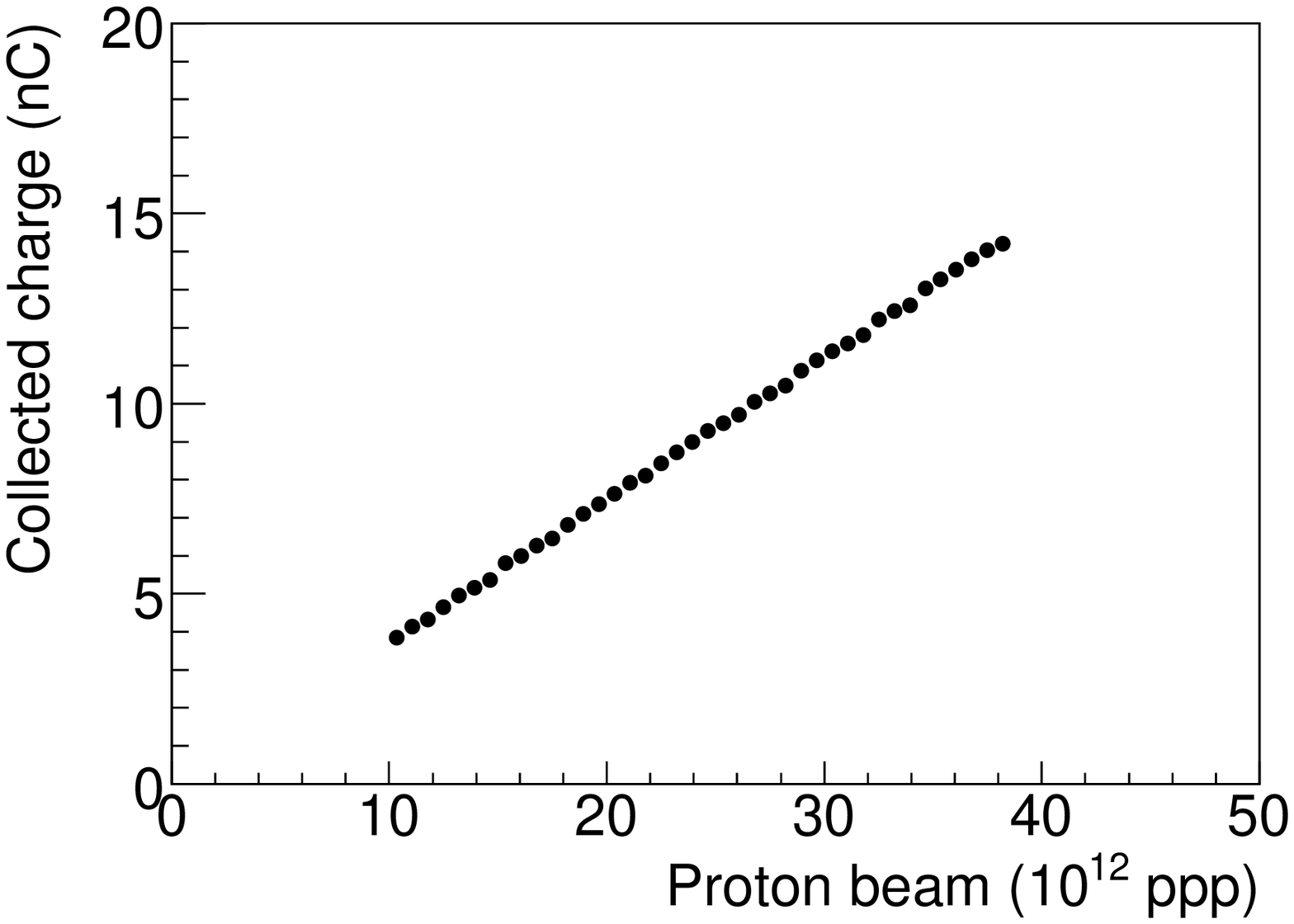}
    \end{center}
  \end{minipage}
  \caption{\label{fig:linearity T968}
  Collected charge to the proton beam intensity for the miniature ionization chamber with the pure He gas (left) and the silicon PIN photodiode (right) measured in the NuMI beamline.}
\end{figure}

In order to measure long-term stabilities of the detectors, the irradiation of the NuMI muon beam to the T2K prototype detectors was continued for half a year.
For the miniature chamber, 200~V was applied.
The results are plotted in Fig.~\ref{fig:stability}, where the collected charges are normalized to the proton beam intensity.
The linearity of the collected charges to the proton beam intensity is within 1\% (Fig.~\ref{fig:linearity T968}).

The gas pressure varied in a range 1.01-1.07~atm.
Therefore, the collected charge of the miniature ionization chamber is divided by the gas pressure to be normalized to equivalent signal at 1~atm.
Since variation of the gas temperature was $\pm0.5^\circ$C around 27.6$^\circ$C, a correction of the temperature is not applied.
Two periods, around January 14th and May 28th, had reduced gas purity\footnote{This was due to an accident during an exchange of the gas cylinders.}, and the efficiency, or the collected charge, jumped up above 1.5~pC/10$^{12}$~ppp due to the Jesse effect.
Except for those periods, the ionization chamber worked stably with no trouble.
The stability is 1.5\% for half a year.

The silicon PIN photodiodes show stable response from beginning to end.
The stability is less than 1\% for half a year.

\subsubsection{Radiation tolerance of the silicon PIN photodiode} \label{sec:silicon radiation}
Three silicon PIN photodiodes were irradiated to the electron beam at around $1\times 10^{9}$~electrons/cm$^2$/pulse for four~hours.
Total fluence was $2.1\times 10^{14}$~electrons/cm$^2$.

After the irradiation of $2.1\times 10^{14}$~electrons/cm$^2$, the collected charge of the silicon PIN photodiodes decreased by approximately 4\%.
However, up to $1\times 10^{14}$~electrons/cm$^2$, no deficit of the collected charge was observed.
Because a \mbox{100-MeV} electron is equivalent to 0.0787~\mbox{1-MeV}~neutrons in terms of the radiation damage~\cite{1MeV neutron}, $1\times 10^{14}$~electrons/cm$^2$ corresponds to the irradiation to the T2K muon beam at 0.75~MW for a month.
This result tells that the photodiode is more tolerant of radiation than the expectation discussed in Sec.~\ref{sec:silicon}.

\subsection{Summary of the detector performance}
The performances of the detectors measured in the beam tests are summarized in Table~\ref{tab:detector performance}.
The applied voltage to the ionization chamber is selected at 200~V so that the response is linear to the beam intensity within 3\%.
The Ar-N$_2$ mixture can be used up to $1.4\times 10^6$~charged~particles/cm$^2$/bunch.
From the linearity and stability, the error on the muon intensity monitoring is expected to be less than 2.9\%.
If the relative gains of the sensors are calibrated with a precision better than 2\%, the error on the relative intensity measurement of each sensor is expected to be 4\%, which satisfies the requirement for the beam direction measurement.

In the T2K experiment, the gas temperature and pressure are kept within 1\% variance; the O$_2$ contamination is reduced down to 2~ppm; adding N$_2$ gas makes the gas response independent of impurities amount; and the two-manifold configuration with the automatic switching regulator reduces the risk of gas impurification during an exchange of the gas cylinders.
This sophisticated gas system ensures a more stable response in T2K run periods than in the beam tests.

The silicon PIN photodiode can be used for a month at 0.75~MW without signal depletion due to the radiation damage.

\begin{table}[t]
 \begin{center}
   \caption{Specification of the muon monitor detectors and their performances measured in the beam tests.}
   \label{tab:detector performance}
   \vspace{\baselineskip}
   \begin{tabular}{|c|c|c|c|}
    \hline
       & \multicolumn{2}{|c|}{Ionization chamber} & \multirow{2}{*}{\parbox{6em}{Silicon PIN photodiode}} \\ \cline{2-3}
       & Ar with 2\% N$_2$ & He with 1\% N$_2$ & \\
    \hline
     Active volume (mm) & \multicolumn{2}{|c|}{$75\times 75\times 3$} & $10\times 10\times 0.3$ \\
    \hline
     Num. of sensors & \multicolumn{2}{|c|}{49} & 49 \\
    \hline
     Intensity (particles/cm$^2$/bunch) & (0.05-1.4)$\times 10^6$ & (0.1-2.3)$\times 10^7$ & $\le 2.2\times 10^7$ \\
    \hline
     Applied voltage (V) & \multicolumn{2}{|c|}{200} & 80 \\
    \hline
     Pulse full width (nsec) & 400 & 700 & $< 500$ \\
    \hline
     Linearity (\%) & 2.4 & 1.7 & 1.9 \\
    \hline
     Long-term stability (\%) & \multicolumn{2}{|c|}{1.5 (pure He)} & $< 1$ \\
    \hline
     Gas temperature ($^\circ$C) & \multicolumn{2}{|c|}{$\sim 34~\pm 0.2$} & \\
    \hline
     Gas pressure (kPa [absolute]) & \multicolumn{2}{|c|}{$130~\pm 0.2$} & \\
    \hline
  \end{tabular}
 \end{center}
\end{table}

\section{Conclusions}

The muon monitor has been developed for the T2K experiment.
It consists of arrays of the ionization chambers and silicon PIN photodiodes.
Based on the beam test results, the muon monitor can monitor the stability of the T2K neutrino beam intensity with a 2.9\% precision and the bunch-by-bunch direction with a 0.25~mrad precision (assuming the uncertainty of the relative sensor gain is less than 2\%) for the expected muon beam intensity of $10^5$-$10^7$~/cm$^2$/bunch.
Every component of the muon monitor is tolerant of radiation.
More stable operation of the ionization chamber than in the beam test is guaranteed by the control system's regulation of the gas pressure, temperature and purity, and by adding N$_2$ gas.
In conclusion, the performance of the muon monitor is expected to satisfy all the requirements in the T2K experiment.

\section{Acknowledgements}

It is a pleasure to thank Prof.~A.~Noda, Dr.~H.~Souda and Mr.~H.~Tongu at Kyoto University and Prof.~T.~Shirai at the National Institute of Radiological Sciences.
They strongly supported us in the electron beam test.
We would like to thank Prof.~S.~Kopp, Mr.~M.~Proga and Dr.~L.~Loiacono at the University of Texas at Austin, Dr.~R.~Zwaska and the staff of Fermilab for the test at the NuMI beamline and for valuable information about an ionization chamber.
We also have had the support and encouragement of the members of the KEK neutrino group.
They have worked together with us for the muon monitor and for the T2K experiment.
This work was supported by MEXT and JSPS with Grant-in-Aid for Scientific Research on Priority Areas 18071007, Young Scientists S 20674004, JSPS Fellows, the Global COE Program ``The Next Generation of Physics, Spun from Universality and Emergence'' and the Japan/U.S. Cooperation Program in the field of High Energy Physics.

\end{document}